\documentclass[12pt, preprint]{aastex}
\usepackage[dvips]{color}
\usepackage{amsmath}
\begin{document}

\title{{\it Herschel} Survey of Galactic OH$^+$, H$_2$O$^+$, and H$_3$O$^+$: Probing the Molecular Hydrogen Fraction and Cosmic-Ray Ionization Rate.\footnote{{\it Herschel} is an ESA space observatory with science instruments provided by European-led Principal Investigator consortia and with important participation from NASA}}

\author{Nick Indriolo\altaffilmark{1,2},
D.~A.~Neufeld\altaffilmark{1},
M.~Gerin\altaffilmark{3},
P.~Schilke\altaffilmark{4},
A.~O.~Benz\altaffilmark{5},
B.~Winkel\altaffilmark{6},
K.~M.~Menten\altaffilmark{6},
E.~T.~Chambers\altaffilmark{4},
John~H.~Black\altaffilmark{7},
S.~Bruderer\altaffilmark{8},
E.~Falgarone\altaffilmark{3},
B.~Godard\altaffilmark{3},
J.~R.~Goicoechea\altaffilmark{9},
H.~Gupta\altaffilmark{10},
D.~C.~Lis\altaffilmark{10,11},
V.~Ossenkopf\altaffilmark{4},
C.~M.~Persson\altaffilmark{7},
P.~Sonnentrucker\altaffilmark{12},
F.~F.~S.~van~der~Tak\altaffilmark{13,14},
E.~F.~van~Dishoeck\altaffilmark{15,8},
Mark~G.~Wolfire\altaffilmark{16},
F.~Wyrowski\altaffilmark{6},
}
\altaffiltext{1}{Department of Physics and Astronomy, Johns Hopkins University, Baltimore, MD 21218, USA}
\altaffiltext{2}{Current address: Department of Astronomy, University of Michigan, Ann Arbor, MI 48109, USA}
\altaffiltext{3}{LERMA, CNRS UMR 8112, Observatoire de Paris \& Ecole Normale Sup\'{e}rieure, Paris, France}
\altaffiltext{4}{I. Physikalisches Institut der Universit\"{a}t zu K\"{o}ln, Z\"{u}lpicher Str. 77, 50937 K\"{o}ln, Germany}
\altaffiltext{5}{Institute of Astronomy, ETH Z\"{u}rich, Switzerland}
\altaffiltext{6}{MPI f\"{u}r Radioastronomie, Bonn, Germany}
\altaffiltext{7}{Department of Earth and Space Sciences, Chalmers University of Technology, Onsala Space Observatory, SE-43992 Onsala, Sweden}
\altaffiltext{8}{Max Planck Institut f\"{u}r Extraterrestrische Physik, Garching, Germany}
\altaffiltext{9}{Instituto de Ciencias de Materiales de Madrid (CSIC), E-28049 Cantoblanco, Madrid, Spain}
\altaffiltext{10}{California Institute of Technology, Pasadena, CA 91125, USA}
\altaffiltext{11}{Sorbonne Universit\'{e}s, Universit\'{e} Pierre et Marie Curie, Paris 6, CNRS, Observatoire de Paris, UMR 8112, LERMA, Paris, France}
\altaffiltext{12}{Space Telescope Science Institute, Baltimore, MD 21218}
\altaffiltext{13}{SRON Netherlands Institute for Space Research, Landleven 12, 9747 AD Groningen, The Netherlands}
\altaffiltext{14}{Kapteyn Astronomical Institute, University of Groningen, The Netherlands}
\altaffiltext{15}{Leiden Observatory, Leiden University, P.O. Box 9513, 2300 RA Leiden, The Netherlands}
\altaffiltext{16}{Department of Astronomy, University of Maryland, College Park, MD 20742}

\begin{abstract}
In diffuse interstellar clouds the chemistry that leads to the formation of the oxygen bearing ions OH$^+$, H$_2$O$^+$, and H$_3$O$^+$ begins with the ionization of atomic hydrogen by cosmic rays, and continues through subsequent hydrogen abstraction reactions involving H$_2$.  Given these reaction pathways, the observed abundances of these molecules are useful in constraining both the total cosmic-ray ionization rate of atomic hydrogen ($\zeta_{\rm H}$) and molecular hydrogen fraction ($f_{\rm H_2}$).
We present observations targeting transitions of OH$^+$, H$_2$O$^+$, and H$_3$O$^+$ made with the {\it Herschel Space Observatory} along 20 Galactic sight lines toward bright submillimeter continuum sources.  Both OH$^+$ and H$_2$O$^+$ are detected in absorption in multiple velocity components along every sight line, but H$_3$O$^+$ is only detected along 7 sight lines. From the molecular abundances we compute $f_{\rm H_2}$ in multiple distinct components along each line of sight, and find a Gaussian distribution with mean and standard deviation $0.042\pm0.018$.  This confirms previous findings that OH$^+$ and H$_2$O$^+$ primarily reside in gas with low H$_2$ fractions.  We also infer $\zeta_{\rm H}$ throughout our sample, and find a log-normal distribution with mean $\log(\zeta_{\rm H})=-15.75$, ($\zeta_{\rm H}=1.78\times10^{-16}$~s$^{-1}$), and standard deviation 0.29 for gas within the Galactic disk, but outside of the Galactic center.  This is in good agreement with the mean and distribution of cosmic-ray ionization rates previously inferred from H$_3^+$ observations.  Ionization rates in the Galactic center tend to be 10--100 times larger than found in the Galactic disk, also in accord with prior studies.
\end{abstract}

\section{INTRODUCTION}

Astrochemistry is a flourishing field, with over 180 molecules (300 when accounting for isotopologues) detected in interstellar and circumstellar environments \citep{lovas2014}. Several of the more recent detections, including those of OH$^+$ \citep{wyrowski2010ohp} and H$_2$O$^+$ \citep{ossenkopf2010}, were made possible as new technology has pushed both ground and space-based observatories into the THz frequency range. Of particular importance was the {\it Herschel Space Observatory} \citep{pilbratt2010}, which offered a view of the THz regime unimpeded by atmospheric absorption. As the inventory of interstellar molecules and complexity of chemical reaction networks grow, it remains imperative that we are able to select the most important reactions governing the abundance of a particular species, and understand how observations of closely related species can be utilized to infer properties of the interstellar medium (ISM).

A basic understanding of how the chemistry involving different species proceeds in the ISM can be garnered from knowledge of a few key atomic and molecular properties, one of which is the first ionization potential (FIP). Neutral-neutral reactions proceed slowly at the relatively low temperatures in diffuse clouds, while ion-neutral reactions are typically much faster and so dominate diffuse cloud chemistry. This necessitates an external ionization mechanism to drive the reaction network.  Species with FIP less than 13.6~eV (below the ionization potential of atomic hydrogen) can be photoionized by far-ultraviolet photons from the interstellar radiation field, and will predominantly be in ionized form.  For species with FIP above 13.6~eV, atomic hydrogen effectively absorbs the ionizing interstellar radiation field, and they remain predominantly in neutral form. Reaction networks of such species are generally initiated by reactions with the ions H$^+$ and H$_3^+$---both of which are primarily formed via cosmic-ray ionization of H and H$_2$, respectively---and so the chemistry surrounding these species can be considered cosmic-ray driven.  Oxygen falls into this latter category ($\mathrm{FIP}=13.62$~eV), so the abundances of various oxygen-bearing molecules are closely linked to the cosmic-ray ionization rate.

Another controlling parameter is the bond-dissociation energy, $D_0$. If $D_0>4.48$~eV (dissociation energy of H$_2$) for a species XH$^+$, then the reaction $\mathrm{X^{+}+H_2\rightarrow XH^{+}+H}$ is exothermic. This is especially important for interstellar chemistry at low temperatures, where there is little kinetic energy to aid in reactions. Dissociation energies of OH$^+$, H$_2$O$^+$, and H$_3$O$^+$ are all greater than 4.48~eV, and O$^+$, OH$^+$, and H$_2$O$^+$ all react exothermically with H$_2$.  As H$_2$ is the most abundant molecule in the universe, the abundances of these molecular ions---specifically with respect to each other---are highly dependent on the amount of H$_2$ available for reactions.

The properties of O and oxygen-bearing ions described above explain the particular utility of OH$^+$, H$_2$O$^+$, and H$_3$O$^+$ in constraining conditions in the ISM.  The formation of each larger molecule requires one more hydrogen abstraction reaction with H$_2$, a process that competes primarily with dissociative recombination with electrons in destroying these ions. This makes the ratios $n({\rm H_2O^+})/n({\rm H_3O^+})$ and $n({\rm OH^+})/n({\rm H_2O^+})$ sensitive to the ratio $n(e)/n({\rm H_2})$.  If the fractional abundance of electrons with respect to total hydrogen ($x_e\equiv n(e)/n_{\rm H}$, where $n_{\rm H}\equiv n({\rm H})+2n({\rm H_2})$)
is known, then these ratios can also be used to infer the molecular hydrogen fraction, $f_{\rm H_2}\equiv 2n({\rm H_2})/n_{\rm H}$. Initial results from observations of OH$^+$ and H$_2$O$^+$ along the sight lines toward W49N and W31C showed $f_{\rm H_2}\lesssim0.1$, implying that both species reside in gas that is primarily atomic \citep{gerin2010,neufeld2010}.  This conclusion is supported by the distribution of OH$^+$ and H$_2$O$^+$ absorption in velocity space, which more closely matches that of atomic H than that of H$_2$O and HF (both tracers of molecular gas).  Similar results are found from observations of electronic transitions of OH$^+$ in the ultraviolet, as it is better correlated with CH$^+$ than with species tracing denser molecular gas such as CH, CN, and OH \citep{krelowski2010,porras2014}. In many sight lines, absorption of OH$^+$ and H$_2$O$^+$ arises at or near the systemic velocity of the background source as well, and is thought to trace the irradiated outflows near massive protostars. Even for these objects though, the OH$^+$/H$_2$O$^+$ and H$_2$O/H$_2$O$^+$ ratios are interpreted as indicating relatively low-density, mostly atomic gas \citep{benz2010,bruderer2010,wyrowski2010}.  Only rarely have OH$^+$ and H$_2$O$^+$ column densities required high molecular fractions \citep[e.g., Orion KL;][]{Gupta2010}.

As the formation of OH$^+$ in diffuse gas begins with the ionization of H by cosmic rays, its abundance is useful in constraining the cosmic-ray ionization rate of atomic hydrogen, $\zeta_{\rm H}$. While other molecules are also used for this purpose, OH$^+$ is unique in its ability to probe $\zeta_{\rm H}$ in gas with $0.01\lesssim f_{\rm H_2}\lesssim0.1$. Estimates of the cosmic-ray ionization rate in diffuse clouds based on molecular abundances have been made for roughly 40 years now, with the earliest utilizing observations of OH and HD in diffuse clouds \citep{odonnell1974,black1977,black1978,hartquist1978}. Those studies typically found ionization rates on the order of a few times 10$^{-17}$~s$^{-1}$, as did later studies using the same molecules \citep{federman1996}, although \citet{vandishoeck1986} required ionization rates of a few times 10$^{-16}$~s$^{-1}$ to reproduce observed column densities with a more detailed model. Findings were generally in good agreement with estimates of $\zeta_{\rm H}$ based on the local interstellar proton spectrum measured by {\it Voyager} \citep{webber1998}.  As a result, it was thought that the cosmic-ray ionization rate was relatively uniform throughout the Galaxy, and a canonical value of $\zeta_{\rm H}=3\times10^{-17}$~s$^{-1}$ was frequently adopted.  

The detection of H$_3^+$ in the ISM \citep{geballe1996} introduced a new, less complicated tracer of the ionization rate, and subsequent surveys of H$_3^+$ pointed to an ionization rate in diffuse clouds nearly ten times larger than that found previously: $\zeta_{\rm H}\approx2\times10^{-16}$~s$^{-1}$ \citep{mccall2003,indriolo2007,indriolo2012}. In addition, the distribution of ionization rates inferred from H$_3^+$ was found to vary by over 1 order of magnitude, suggesting that the low-energy cosmic-ray flux is not uniform throughout the Galaxy.  It now seems likely that most early estimates of $\zeta_{\rm H}$ were too low because they assumed that nearly every instance of hydrogen being ionized by a cosmic ray led to the formation of OH or HD.  However, destruction of H$^+$ by polycyclic aromatic hydrocarbons (PAHs) and small grains is highly competitive with the charge transfer reactions driving the oxygen and deuterium chemistries \citep{wolfire2003}, making the chemical pathways from H$^+$ to OH and HD ``leaky.''  This mechanism was recognized by \citet{liszt2003} as a way to reconcile the differences in ionization rates inferred from OH and HD with those inferred from H$_3^+$.  Neutralization of H$^+$ on grains is also important in the chemistry leading to OH$^+$ and H$_2$O$^+$, and its effects are now accounted for when using these species to infer the ionization rate \citep{neufeld2010,hollenbach2012,indriolo2012w51}.

While infrared and radio observations of interstellar molecules---carefully interpreted in the context of astrochemical models---can be used to determine the density of low-energy cosmic rays ($E\lesssim10$~MeV), gamma-ray observations provide a complementary probe of high-energy cosmic rays ($E\gtrsim300$~MeV).  The latter interact with atomic nuclei in the interstellar gas, producing neutral pions ($\pi^0$) that rapidly decay into pairs of gamma-ray photons \citep{beringer2012}. Observations of these gamma-rays can be used to estimate the density of high-energy cosmic rays as a function of location within the Galaxy. Our understanding of the gamma-ray sky has greatly improved following the launch of the {\it Fermi Gamma-ray Space Telescope}, with recent observations of the outer Galaxy suggesting that the cosmic-ray density is relatively uniform outside the solar circle,
and declines less rapidly with Galactocentric radius ($R_{\rm gal}$) than predicted by propagation models \citep{ackermann2011}.  An interesting question is whether the density of low-energy particles shows the same behavior, or whether the significantly smaller amount of material through which such particles can travel before losing all of their energy leads to a different result.

Observations of H$_3^+$ have primarily been limited to the local ISM \citep[within about 2~kpc of the Sun;][]{mccall2002,indriolo2012} due to the necessity for high spectral resolution and high continuum level signal-to-noise ratio (S/N).  The most notable exceptions have been ongoing surveys of the Galactic center region which reveal a large amount of warm, diffuse gas that experiences a large flux of cosmic rays, with ionization rates above $10^{-15}$~s$^{-1}$ \citep{oka2005,goto2008,goto2011,geballe2010}.  Even the dense gas in the Galactic center experiences a cosmic-ray ionization rate 10--100 times larger than the dense gas elsewhere in the Galactic disk, as determined from observations of H$_3$O$^+$, H$^{13}$CO$^+$, and H$_3^+$ \citep{vandertak2000,vandertak2006,goto2013_GC,goto2014}, suggesting an increased particle flux in the Galactic center at all energies.  Still, all of these observations have only probed ionization rates in the Galactic center and the local ISM. To expand this coverage to wider portions of the Galaxy and answer the question posed above, other tracers of the cosmic-ray ionization rate are needed, and {\it Herschel} provided the opportunity to use observations of OH$^+$ and H$_2$O$^+$ for this purpose. 

\subsection{Oxygen Chemistry}

Oxygen chemistry in diffuse clouds is thought to be relatively simple \citep[e.g.,][]{hollenbach2012}, with the network of ion-neutral reactions initiated by the ionization of atomic hydrogen by cosmic rays,
\begin{equation}
{\rm H} + {\rm CR} \rightarrow {\rm H}^+ + e^- + {\rm CR'}.
\label{reac_CR_H}
\end{equation} 
Ionization of H is followed by endothermic charge transfer to oxygen to form O$^+$,
\begin{equation}
{\rm H}^+ + {\rm O} + \Delta E \longleftrightarrow {\rm O}^+ + {\rm H},
\label{reac_Hp_Op}
\end{equation}
where $\Delta E=226$~K represents the endothermicity of the forward reaction (for O in the lowest energy fine-structure level, $^3P_2$, of the ground state), and the double-sided arrow shows that the exothermic back-reaction proceeds uninhibited.  The rate of the forward reaction for oxygen in each of the $^3P_J$ ($J=0,1,2$) fine-structure levels is highly dependent on the gas kinetic temperature (about 100~K on average in diffuse clouds), and the total forward rate on the relative population in the fine-structure levels of atomic oxygen \citep{stancil1999}.\footnote{Rate coefficients for reaction (\ref{reac_Hp_Op}) at low temperature are based solely on quantum mechanical calculations and remain uncertain. It is possible that the most frequently adopted coefficients \citep{stancil1999} are too large \citep{spirko2003}, in which case the oxygen chemistry proceeds more slowly.  This may contribute to the low efficiency in forming OH$^+$ from H$^+$ discussed below.}  Also, the O in reaction (\ref{reac_Hp_Op}) competes with electrons and neutral and charged small grains and PAHs in destroying H$^+$ \citep{wolfire2003},
\begin{align}
{\rm H}^+ + {\rm PAH} & \rightarrow {\rm H} + {\rm PAH}^+, \nonumber \\
{\rm H}^+ + {\rm PAH}^- & \rightarrow {\rm H} + {\rm PAH}, \nonumber \\
{\rm H}^+ + e^- & \rightarrow {\rm H} + h\nu, \nonumber
\label{reac_Hp_remove}
\end{align}
all of which decrease the efficiency at which ionization of H leads to the formation of OH$^+$ \citep{liszt2003}.  Once O$^+$ is formed it can undergo the back-reaction with H, or it can react with H$_2$ to form OH$^+$,
\begin{equation}
\mathrm{O^+ + H_2 \rightarrow OH^+ + H},
\label{reac_Op_H2}
\end{equation}
which is either destroyed by further hydrogen abstraction to form H$_2$O$^+$,
\begin{equation}
\mathrm{OH^{+} + H_{2} \rightarrow H_{2}O^{+} + H},
\label{reac_OH+_H2}
\end{equation}
or by dissociative recombination with electrons,
\begin{equation}
\mathrm{OH^{+}} + e^{-} \rightarrow \mathrm{products}.
\label{reac_OH+_e}
\end{equation}
The same is true for H$_2$O$^+$,
\begin{equation}
\mathrm{H_{2}O^{+} + H_{2} \rightarrow H_{3}O^{+} + H},
\label{reac_H2O+_H2}
\end{equation}
\begin{equation}
\mathrm{H_{2}O^{+}} + e^{-} \rightarrow \mathrm{products},
\label{reac_H2O+_e}
\end{equation}
but H$_3$O$^+$ is primarily destroyed by dissociative recombination with electrons,
\begin{equation}
\mathrm{H_{3}O^{+}} + e^{-} \rightarrow \mathrm{products},
\label{reac_H3O+_e}
\end{equation}
as further hydrogen abstraction reactions with H$_2$ do not proceed. It is apparent from reactions (\ref{reac_Op_H2}) through (\ref{reac_H3O+_e}) that the abundances of these species are controlled by competition between hydrogen abstraction from H$_2$ and dissociative recombination with electrons.

In addition to reaction (\ref{reac_Op_H2}), it is possible for the oxygen chemistry to be driven by the reaction
\begin{equation}
{\rm O} + {\rm H_3^+} \rightarrow {\rm OH}^+ + {\rm H_2},
\label{reac_O_H3+}
\end{equation}
where H$_3^+$ is formed following cosmic-ray ionization of H$_2$ and subsequent reaction of H$_2^+$ with another H$_2$.  To compete with reactions (\ref{reac_CR_H})--(\ref{reac_Op_H2}), this pathway requires a substantial fraction of hydrogen to be in molecular form.  In gas with small $f_{\rm H_2}$, cosmic-ray ionization will produce significantly more H$^+$ than H$_2^+$.  Additionally, H$_2^+$ is likely to undergo charge exchange with the abundant H (i.e., $\rm{H}_2^+ + {\rm H}\rightarrow{\rm H}_2 + {\rm H}^+$), prior to finding another H$_2$, limiting the formation of H$_3^+$.  Combined, these two effects inhibit the pathway to OH$^+$ through reaction (\ref{reac_O_H3+}) in gas that is mostly atomic. As we will show that most of the gas under consideration in this study is diffuse with low molecular hydrogen fraction, we omit this formation route from our analysis, and focus instead on the pathway following reactions (\ref{reac_CR_H})--(\ref{reac_Op_H2}).

The utility of OH$^+$ and H$_2$O$^+$ abundances in constraining the molecular hydrogen fraction and cosmic-ray ionization rate has been demonstrated in multiple studies \citep[e.g.,][]{gerin2010,neufeld2010,indriolo2012w51}, and makes observations of these species important for studying properties of the diffuse Galactic ISM.  As part of the PRISMAS (PRobing InterStellar Molecules with Absoprtion line Studies) Key Program, and motivated by the astrochemical and astrophysical considerations discussed above, we carried out a survey of OH$^+$ and H$_2$O$^+$ line absorption toward nine bright submillimeter continuum sources using the Heterodyne Instrument for the Far-Infrared \citep[HIFI;][]{degraauw2010} on {\it Herschel}. The target sources all lie in the Galactic plane, and are all known to exhibit absorption by molecules in foreground molecular clouds not associated with the sources themselves.  Results from three of the targeted sight lines---W31C, W49N, and W51e---have been reported previously, but those studies only utilized a portion of the data that are now available. In this paper we have compiled the full set of observations of OH$^+$ and H$_2$O$^+$ from PRISMAS, as well as observations from other {\it Herschel} programs toward 11 more sight lines with the intent of exploring $f_{\rm H_2}$ and $\zeta_{\rm H}$ throughout the Galaxy.  The sample of observations is described in Section \ref{sec_obs}; the analysis of these data and findings in Section \ref{sec_analysis}; and a discussion of the findings in Section \ref{sec_disc}.

\section{OBSERVATIONS} \label{sec_obs}

\begin{figure}
\epsscale{1.0}
\plotone{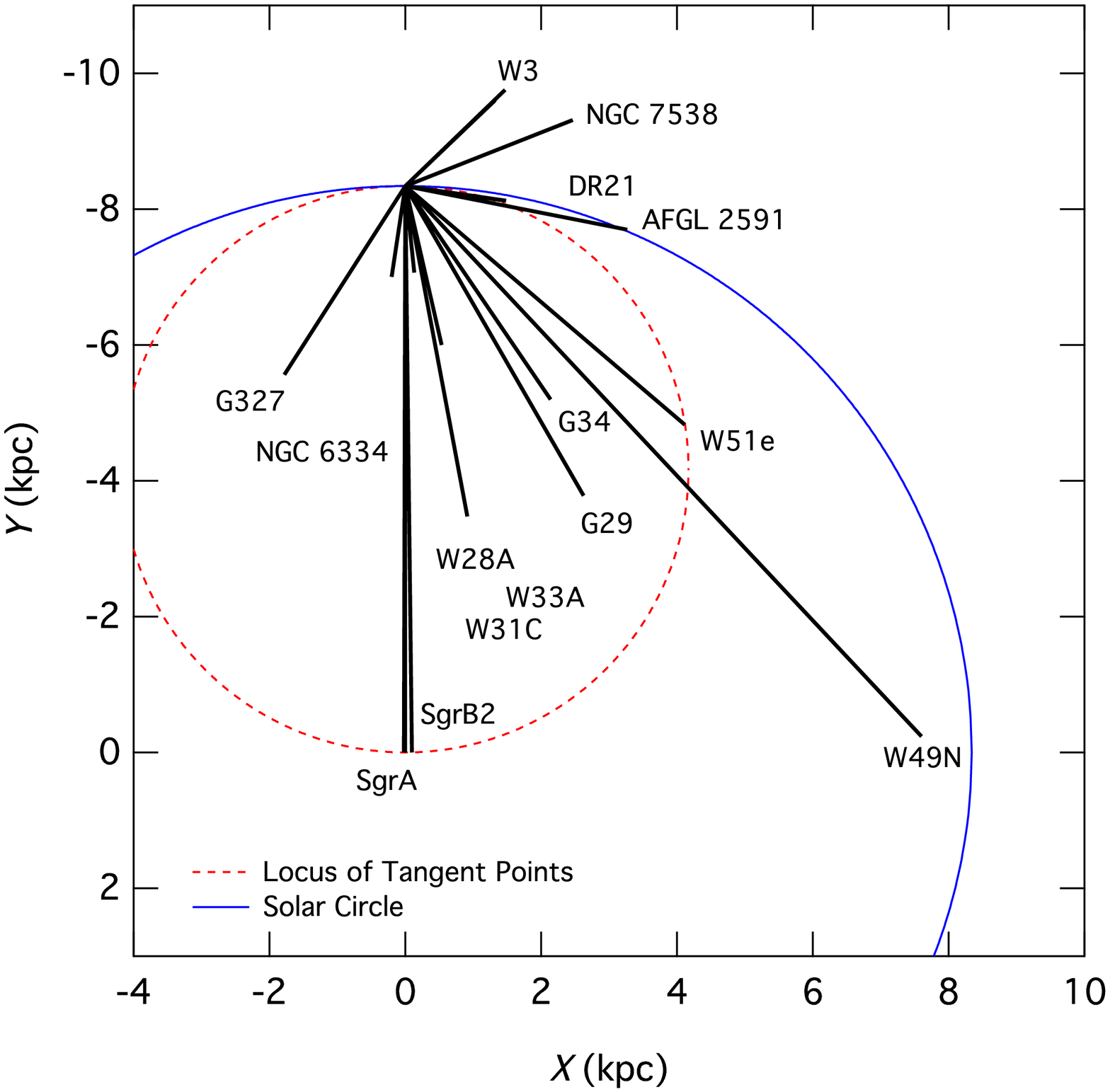}
\caption{Distribution of observed sight lines as viewed from the North Galactic Pole.  Many source names have been shortened for clarity.  The Galactic center is located at (0,0), and the Sun is assumed to be 8.34~kpc away \citep{reid2014}.  The blue solid curve shows the solar circle, and the red dashed curve the locus of tangent velocities.  Only the W49N line of sight significantly samples both near and far kinematic distances, leading to severe blending of absorption features arising in physically separated clouds. }
\label{fig_map_los}
\end{figure}

All observations presented herein were made using the HIFI instrument on board {\it Herschel}.  Multiple transitions of the oxygen-bearing ions OH$^+$, H$_2$O$^+$, and H$_3$O$^+$ were targeted in several different observing programs. A list of the targeted transitions is given in Table \ref{tbl_transitions}. Sight lines along which observations were made are listed in Table \ref{tbl_target}, and Figure \ref{fig_map_los} shows their distribution in the Galactic disk.  Observations were performed using the dual beam switch mode, with the telescope beam centered at the coordinates given in Table \ref{tbl_target}, and the reference positions located at offsets of 3\arcmin\ on either side of each source.  Multiple local oscillator (LO) frequencies separated by small offsets were used to confirm the assignment of any observed spectral feature to either the upper or lower sideband of the double sideband HIFI receivers.  All data were acquired using the Wide Band Spectrometer, which provides a spectral resolution of 1.1~MHz and a bandwidth of $\sim 4$~GHz.   As discussed in \citet{neufeld2010}, the data were processed using the standard HIFI pipeline (versions 9.1 through 11.1 depending on when data were downloaded) to Level 2, providing fully calibrated spectra with the intensities expressed as antenna temperature. The resultant spectra were co-added to recover the signal-to-noise ratio that would have been obtained at a single LO setting.  Spectra obtained for the horizontal and vertical polarizations were found to be very similar in their appearance and noise characteristics and were likewise coadded.

\section{ANALYSIS AND RESULTS} \label{sec_analysis}

\subsection{Spectra}

Table \ref{tbl_obs} lists the double sideband (DSB) continuum antenna temperature, $T_A({\rm DSB})$, measured for each of the target sources at the relevant observing frequencies, together with the root mean square (RMS) noise in the co-added spectra.   Because HIFI employs double sideband receivers, the complete absorption of radiation in any observed spectral line reduces the antenna temperature to roughly one-half its continuum value.  The fractional transmission at any frequency is given by
\begin{equation}
\frac{F(\nu)}{F({\rm cont})}=\left[T_A(\nu)-\frac{T_A({\rm DSB})}{(1+\Gamma)}\right]\left[\frac{T_A({\rm DSB})}{(1+\Gamma^{-1})}\right]^{-1},
\label{eq_transmission}
\end{equation}
where $\Gamma$ is defined as the continuum antenna temperature coming from the sideband containing the frequency of interest divided by the continuum antenna temperature coming from the opposite sideband.  In the special case with $\Gamma=1$, i.e., both sidebands contribute equally to $T_A({\rm DSB})$, equation (\ref{eq_transmission}) simplifies to $F(\nu)/F({\rm cont})=2T_A(\nu)/T_A({\rm DSB}) - 1$.  For all transitions of H$_2$O$^+$ and H$_3$O$^+$, and for the 909~GHz and 1033~GHz transitions of OH$^+$ $\Gamma=1$ is adopted in converting spectra from antenna temperature to fractional transmission \citep[justified by measurements of sideband ratios reported in][]{higgins2014}.  In cases where absorption by the 971~GHz transition of OH$^+$ is saturated, the relative intensities of the different hyperfine components of the transition are assumed constant, and the measured optical depth of the weakest component is used to predict the optical depth of the strongest component. This enables the determination of $\Gamma$, and is an important step as small changes in saturated absorption correspond to large differences in optical depth and thus inferred column density.  

The resulting spectra for all observed transitions and sources are presented in Figures \ref{fig_sgra20_spectra} through \ref{fig_ngc6334In_spectra}.  OH$^+$ and the {\it ortho} spin modification of H$_2$O$^+$ are detected in absorption toward all of the targeted sight lines, while H$_3$O$^+$ and the {\it para} form of H$_2$O$^+$ are each seen in absorption toward only 7 sight lines.  Fits to the absorption features (fitting procedure described below) are shown as red curves (blue curves for the sight lines toward Sgr~B2), and for transitions with hyperfine splitting the green curves show only absorption due to the strongest hyperfine component.  Stick diagrams above spectra mark the hyperfine structure when applicable.  

\subsection{Spectral Fitting}

\begin{figure}
\epsscale{0.8}
\plotone{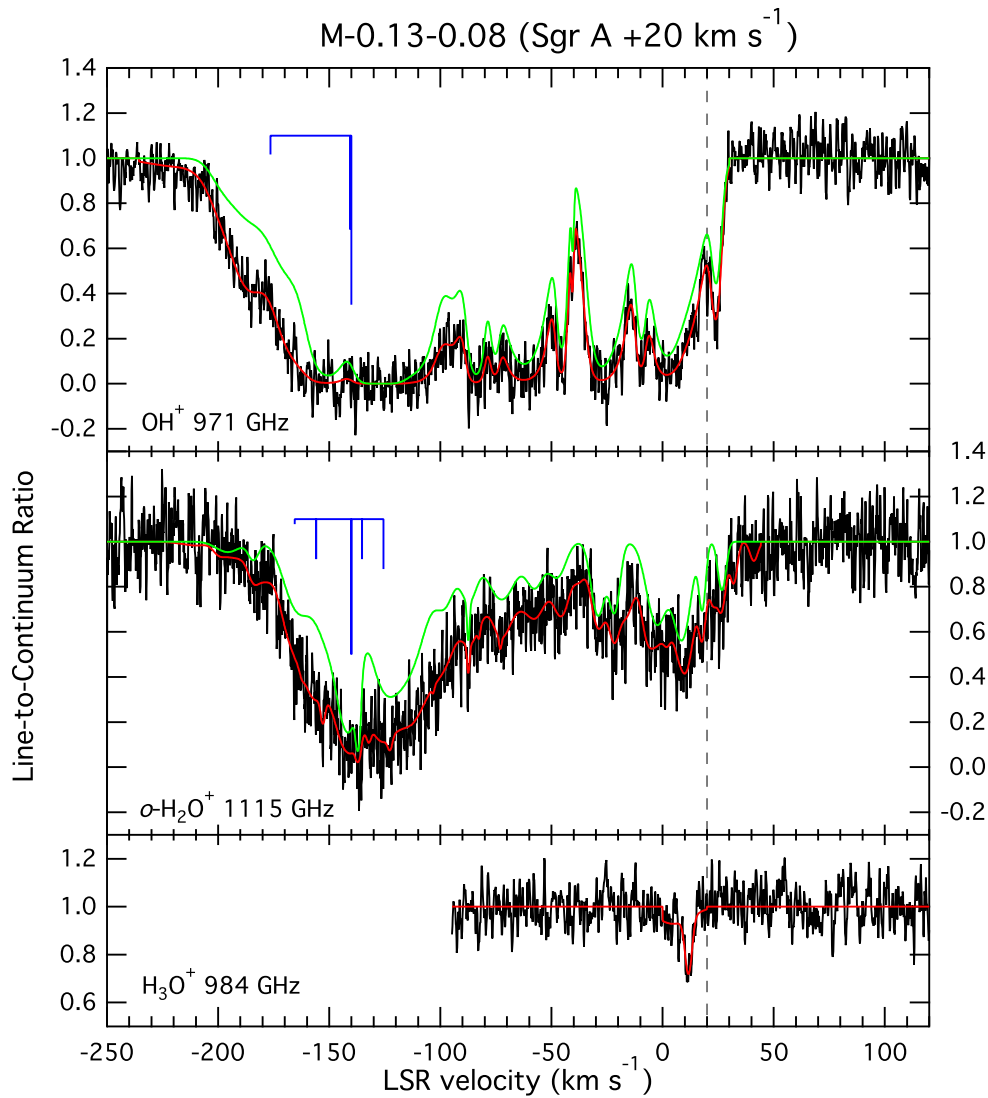}
\caption{Single sideband normalized spectra toward M$-$0.13$-$0.08 (SgrA $+$20 km~s$^{-1}$ cloud) showing transitions of OH$^+$, H$_2$O$^+$, and H$_3$O$^+$.  Stick diagrams above spectra show the hyperfine structure where applicable. Red curves are fits to the absorption features, and green curves show only the strongest hyperfine component of the fits. The vertical dashed line marks the systemic velocity of the background source.  Vertical axes give line-to-continuum ratio, with labels alternating between the left and right sides for clarity.}
\label{fig_sgra20_spectra}
\end{figure}

\begin{figure}
\epsscale{0.8}
\plotone{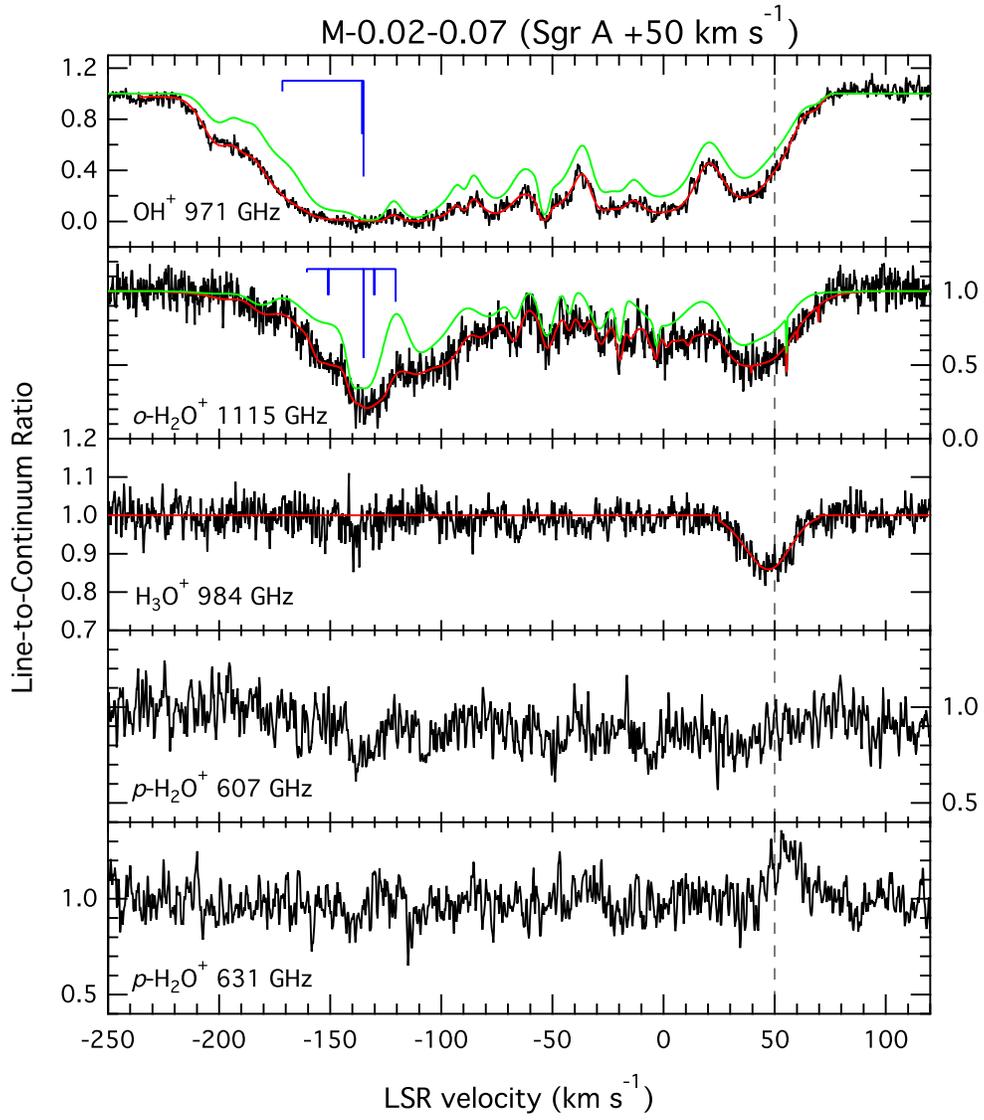}
\caption{Same as Figure \ref{fig_sgra20_spectra} but for M$-$0.02$-$0.07 (SgrA $+$50 km~s$^{-1}$ cloud).}
\label{fig_sgra50_spectra}
\end{figure}

\begin{figure}
\epsscale{0.8}
\plotone{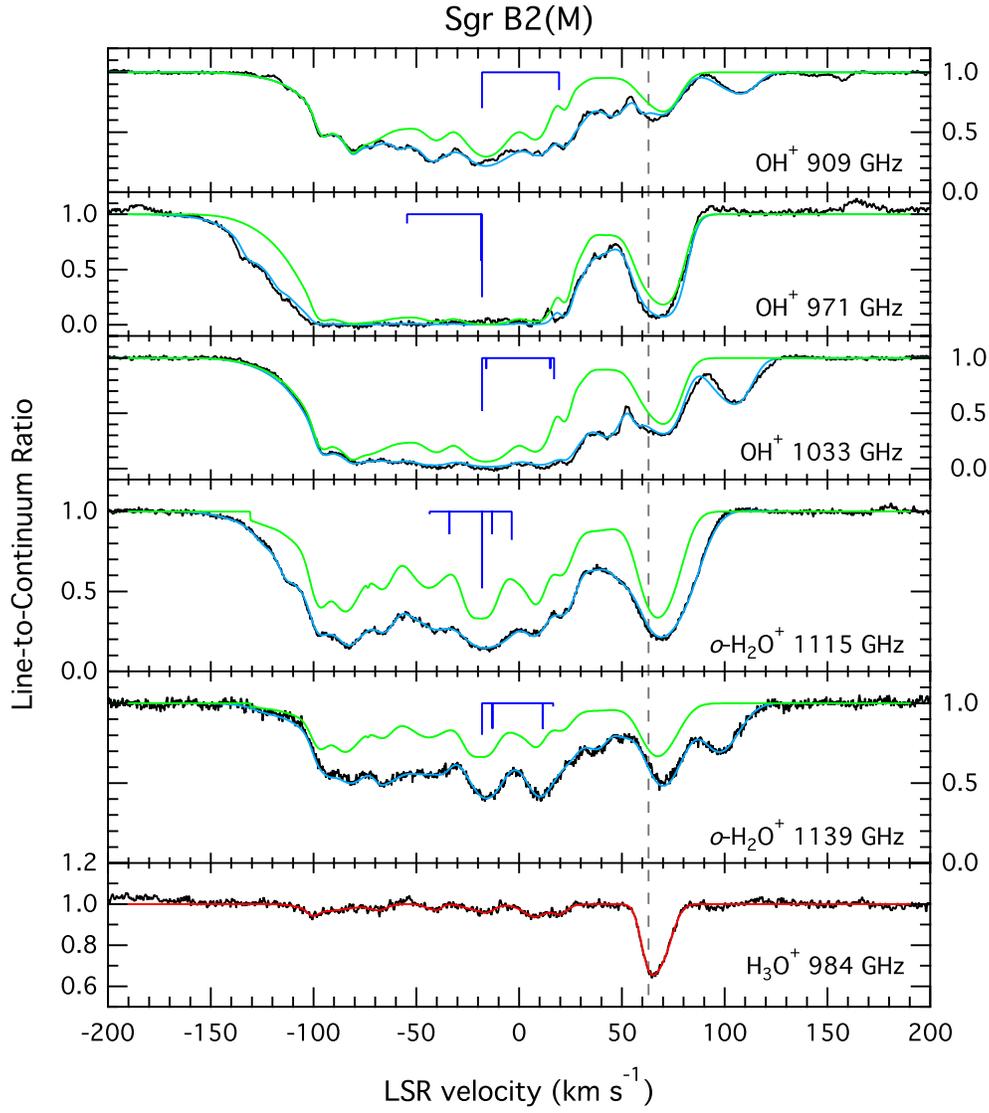}
\caption{Same as Figure \ref{fig_sgra20_spectra} but for Sgr B2(M). In this case, however, fits shown by blue curves were made by using
 absorption from all relevant transitions (e.g., 909~GHz, 971~GHz, and 1033~GHz transitions of OH$^+$) simultaneously.}
\label{fig_sgrb2m_spectra}
\end{figure}

\begin{figure}
\epsscale{0.8}
\plotone{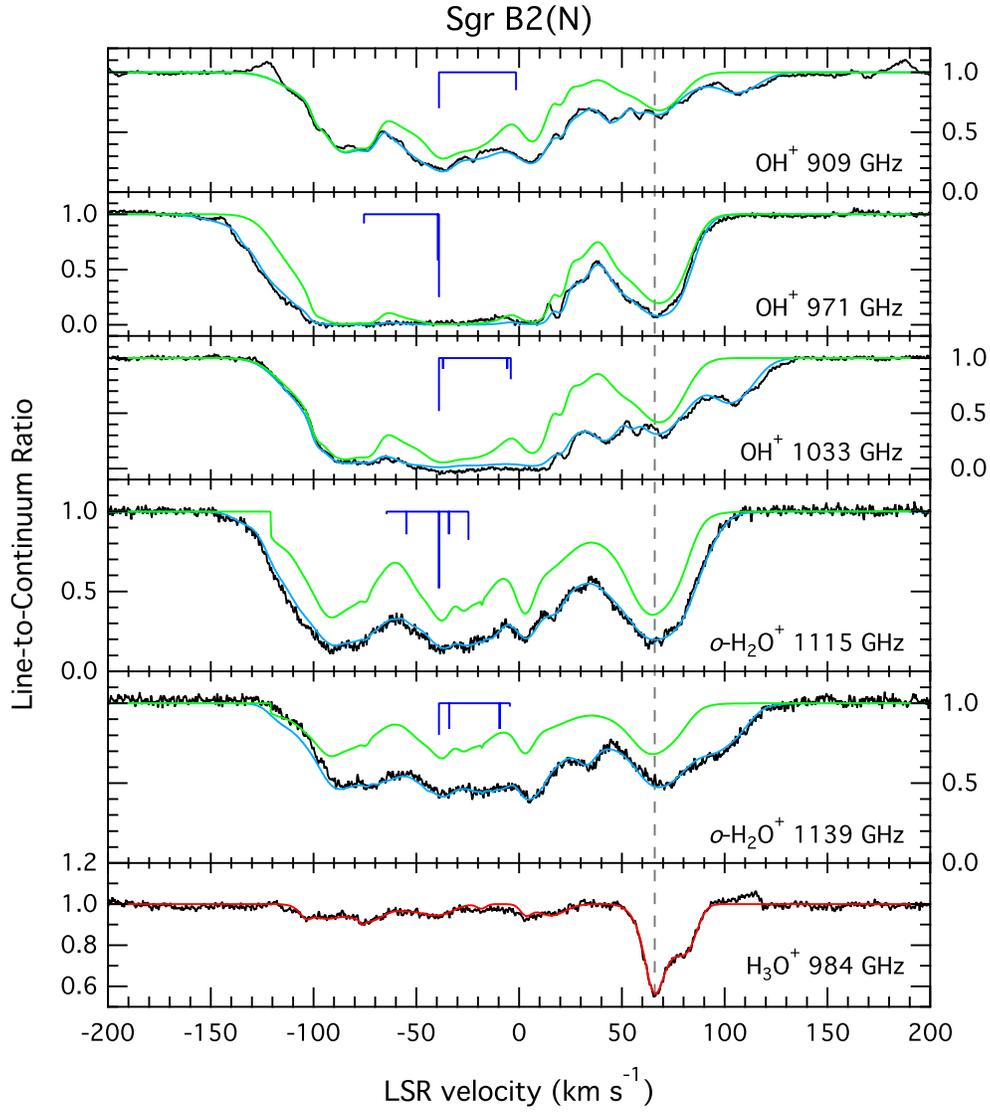}
\caption{Same as Figure \ref{fig_sgrb2m_spectra} but for Sgr B2(N).}
\label{fig_sgrb2n_spectra}
\end{figure}

\begin{figure}
\epsscale{0.8}
\plotone{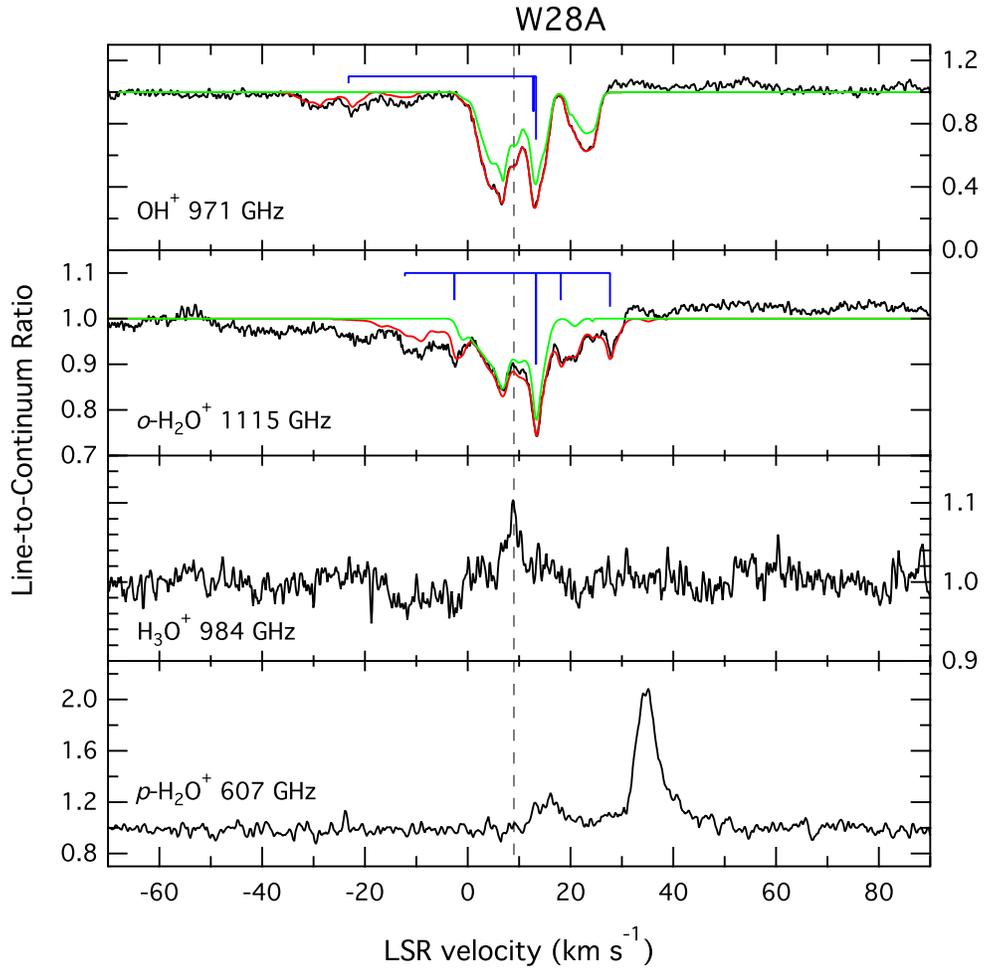}
\caption{Same as Figure \ref{fig_sgra20_spectra} but for W28A.}
\label{fig_w28a_spectra}
\end{figure}

\begin{figure}
\epsscale{0.8}
\plotone{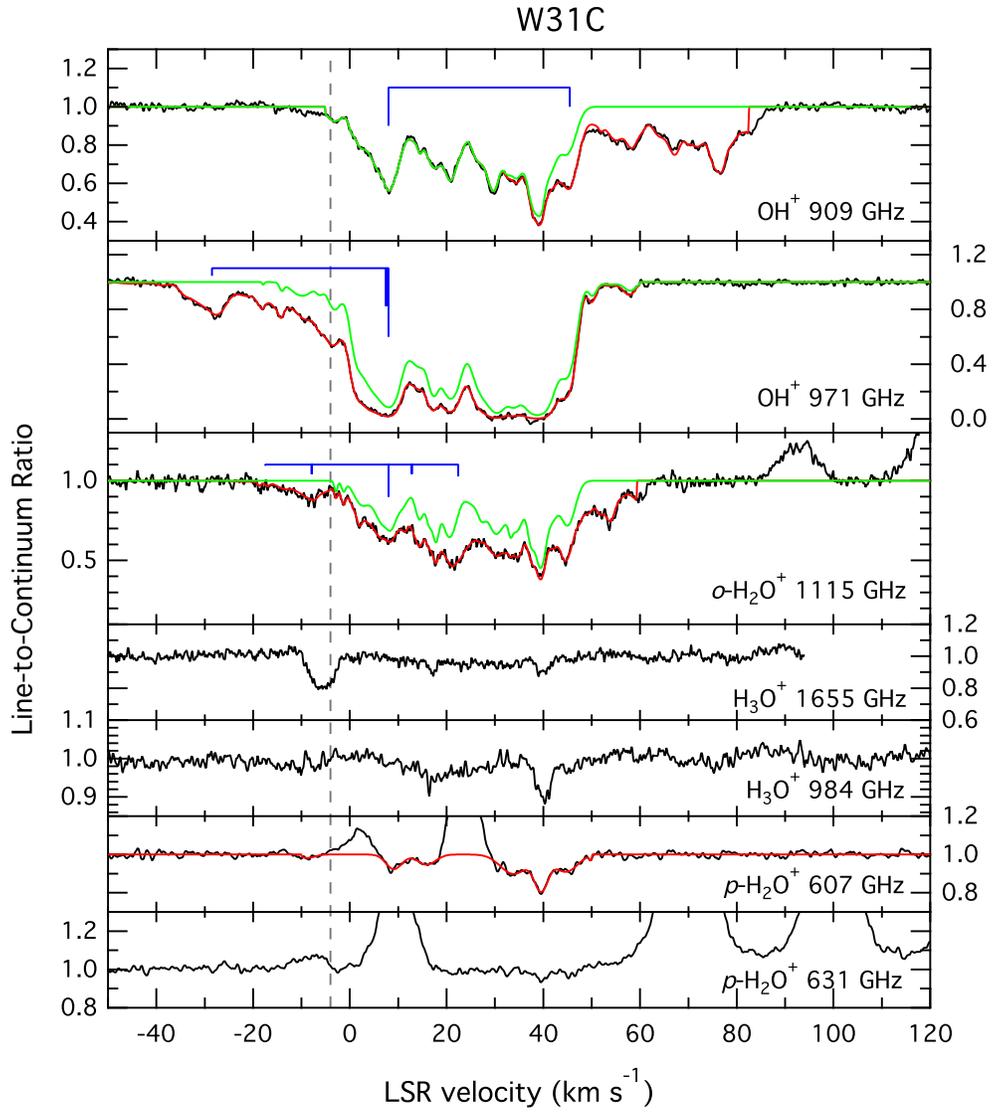}
\caption{Same as Figure \ref{fig_sgra20_spectra} but for W31C.}
\label{fig_w31c_spectra}
\end{figure}

\begin{figure}
\epsscale{0.8}
\plotone{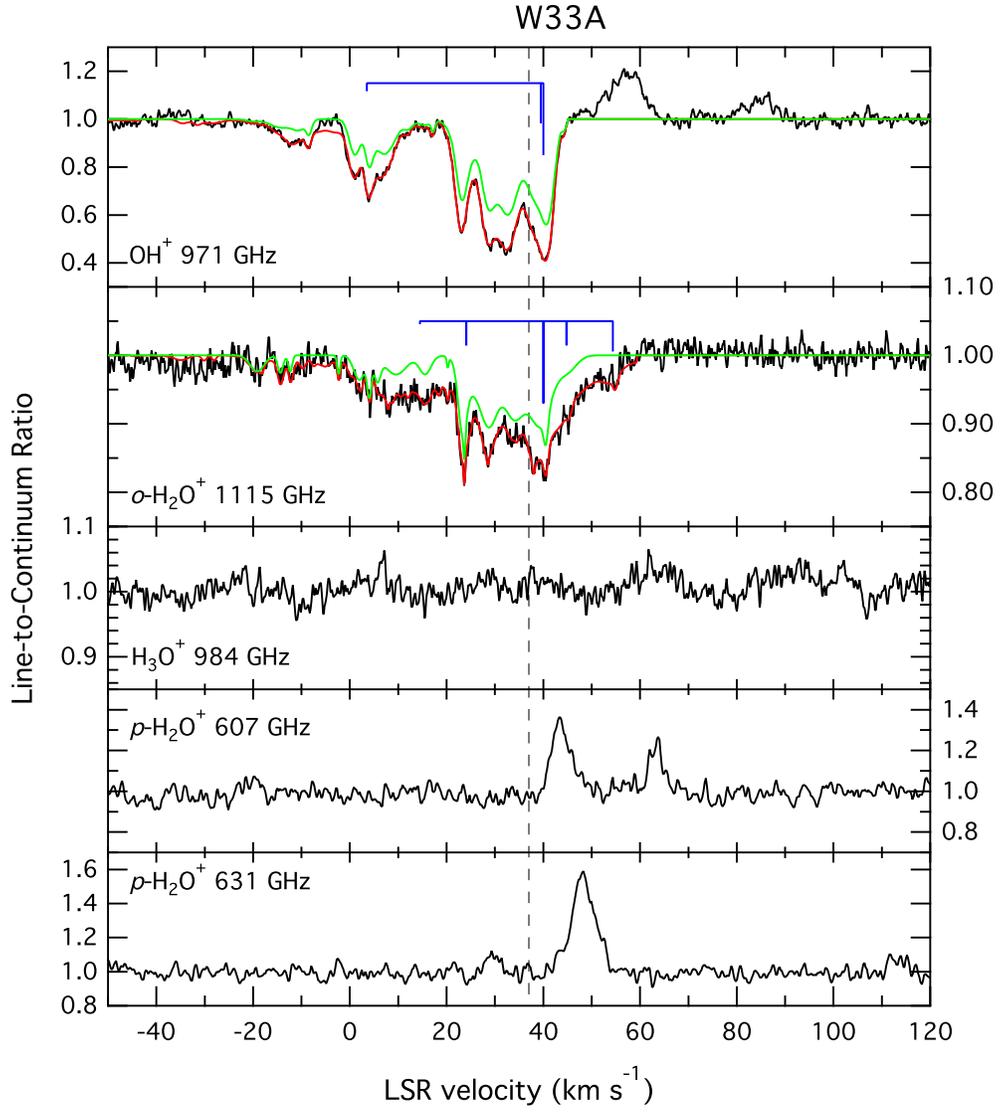}
\caption{Same as Figure \ref{fig_sgra20_spectra} but for W33A.}
\label{fig_w33a_spectra}
\end{figure}

\begin{figure}
\epsscale{0.8}
\plotone{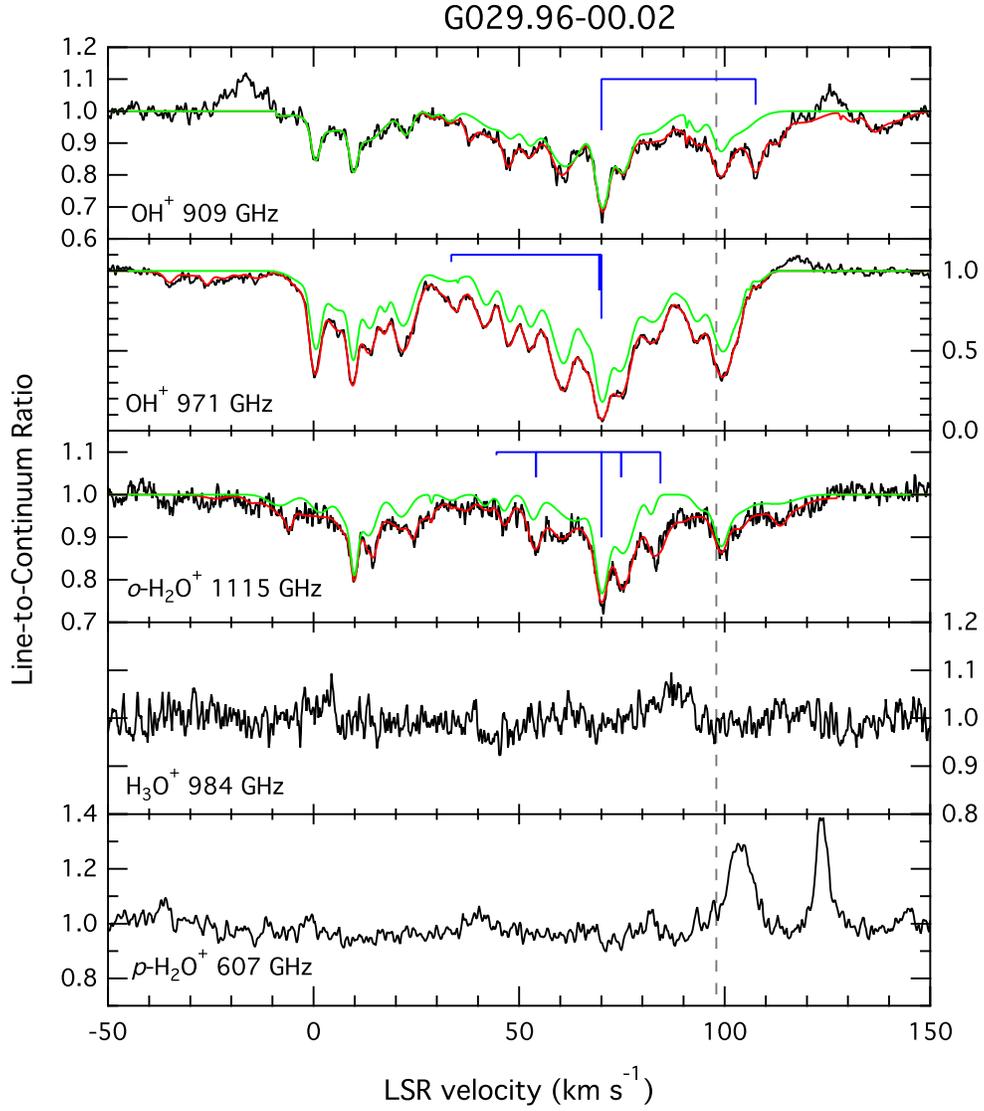}
\caption{Same as Figure \ref{fig_sgra20_spectra} but for G029.96$-$00.02.}
\label{fig_g29_spectra}
\end{figure}

\begin{figure}
\epsscale{0.8}
\plotone{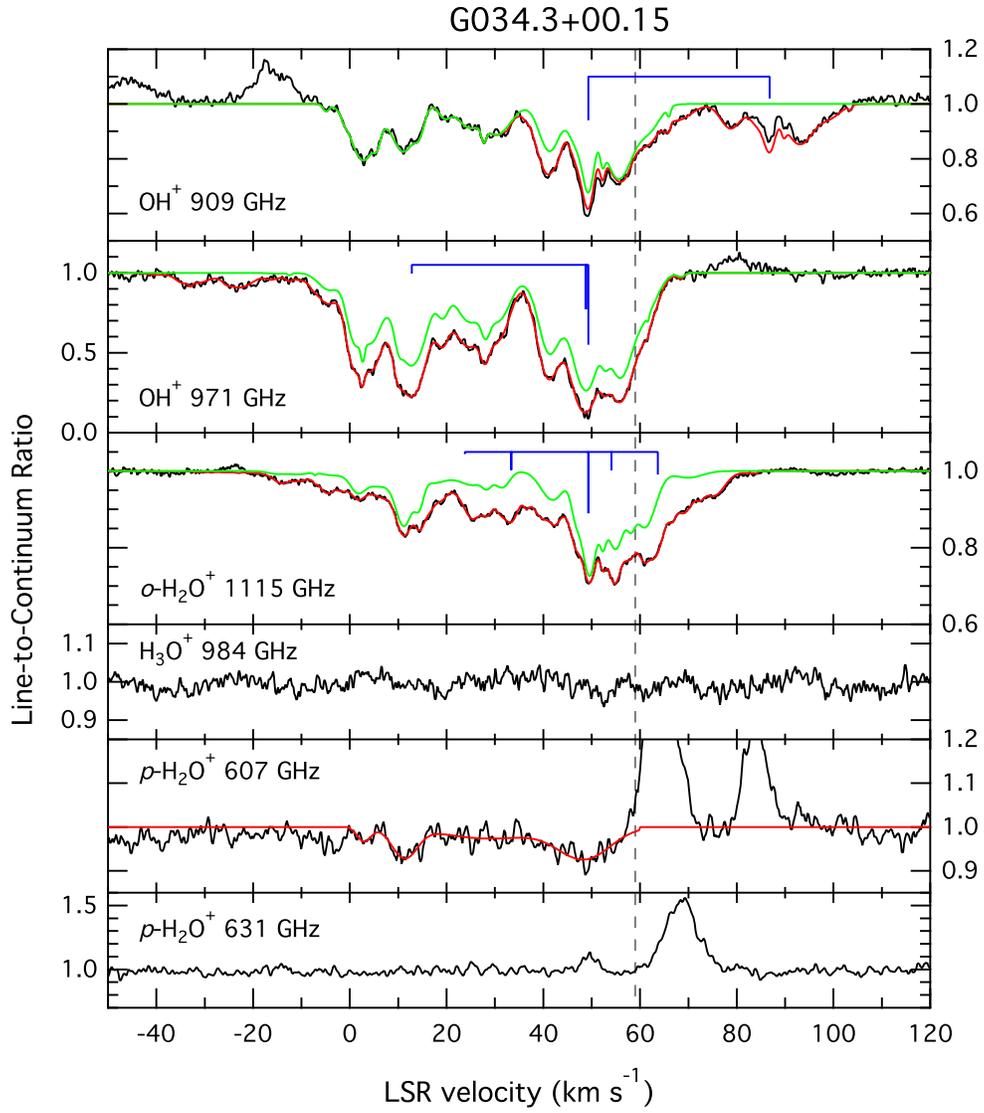}
\caption{Same as Figure \ref{fig_sgra20_spectra} but for G034.3+00.15.}
\label{fig_g34_spectra}
\end{figure}

\begin{figure}
\epsscale{0.8}
\plotone{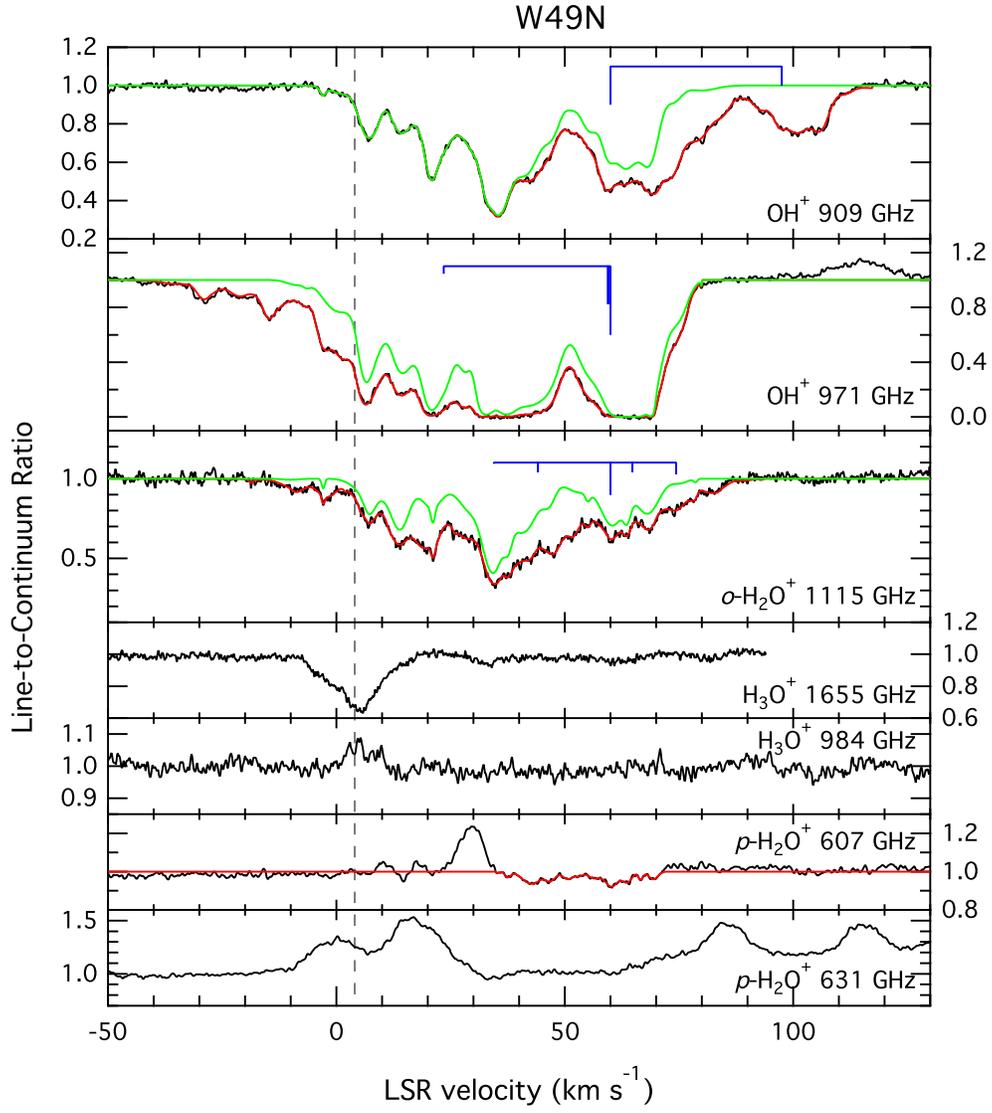}
\caption{Same as Figure \ref{fig_sgra20_spectra} but for W49N.}
\label{fig_w49n_spectra}
\end{figure}

\begin{figure}
\epsscale{0.8}
\plotone{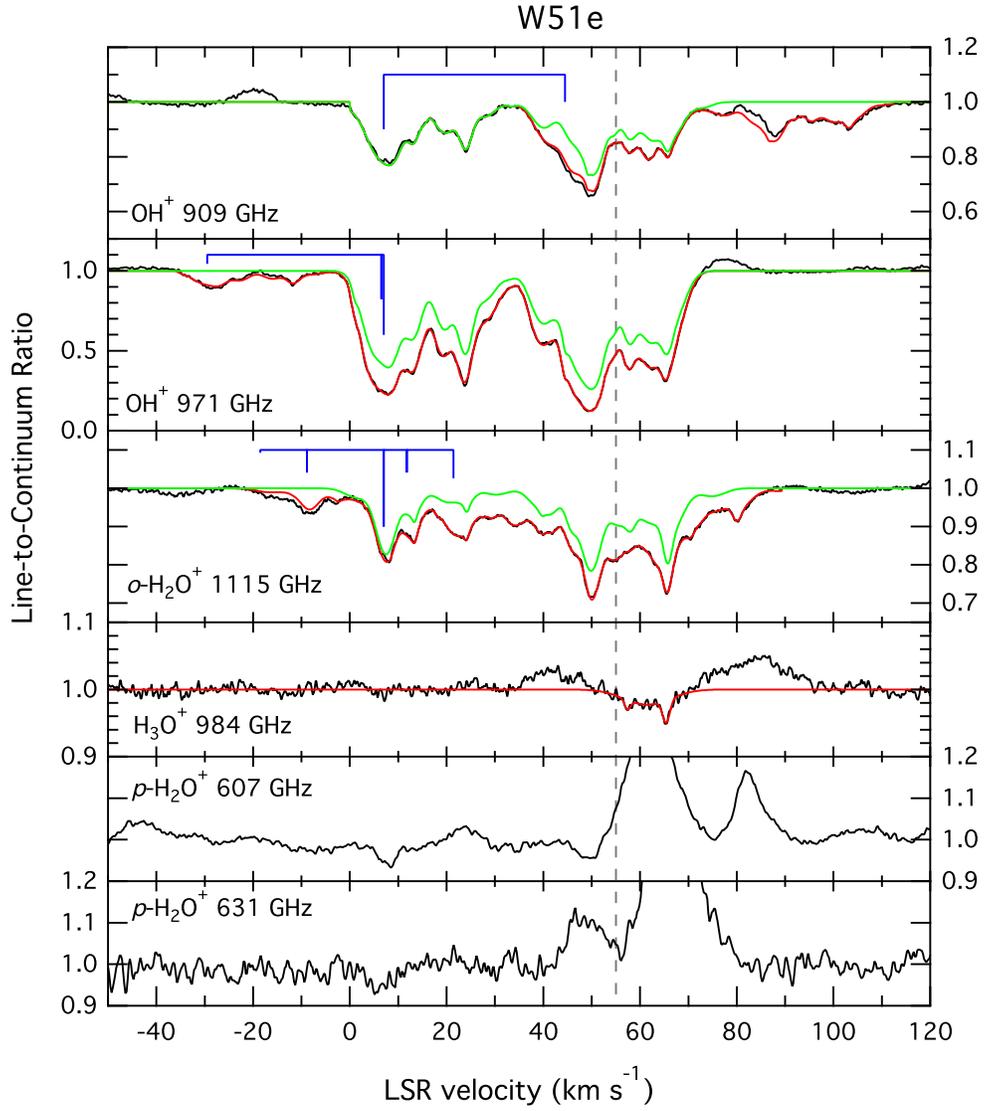}
\caption{Same as Figure \ref{fig_sgra20_spectra} but for W51e.}
\label{fig_w51_spectra}
\end{figure}

\begin{figure}
\epsscale{0.8}
\plotone{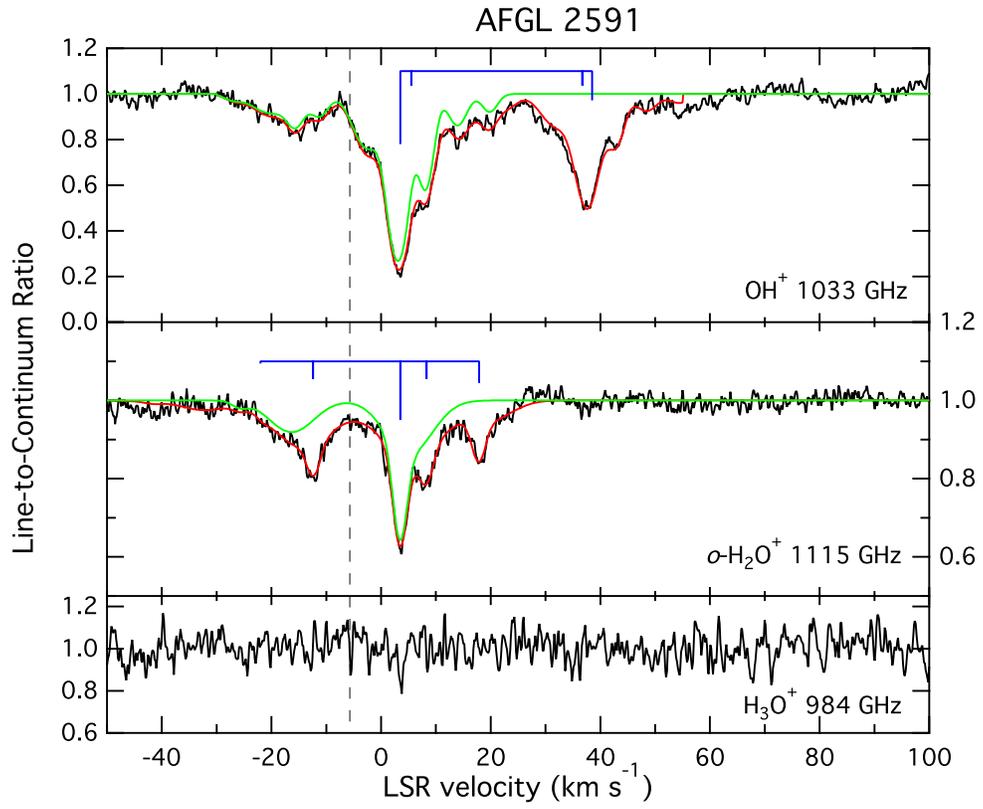}
\caption{Same as Figure \ref{fig_sgra20_spectra} but for AFGL 2591.}
\label{fig_afgl2591_spectra}
\end{figure}

\begin{figure}
\epsscale{0.8}
\plotone{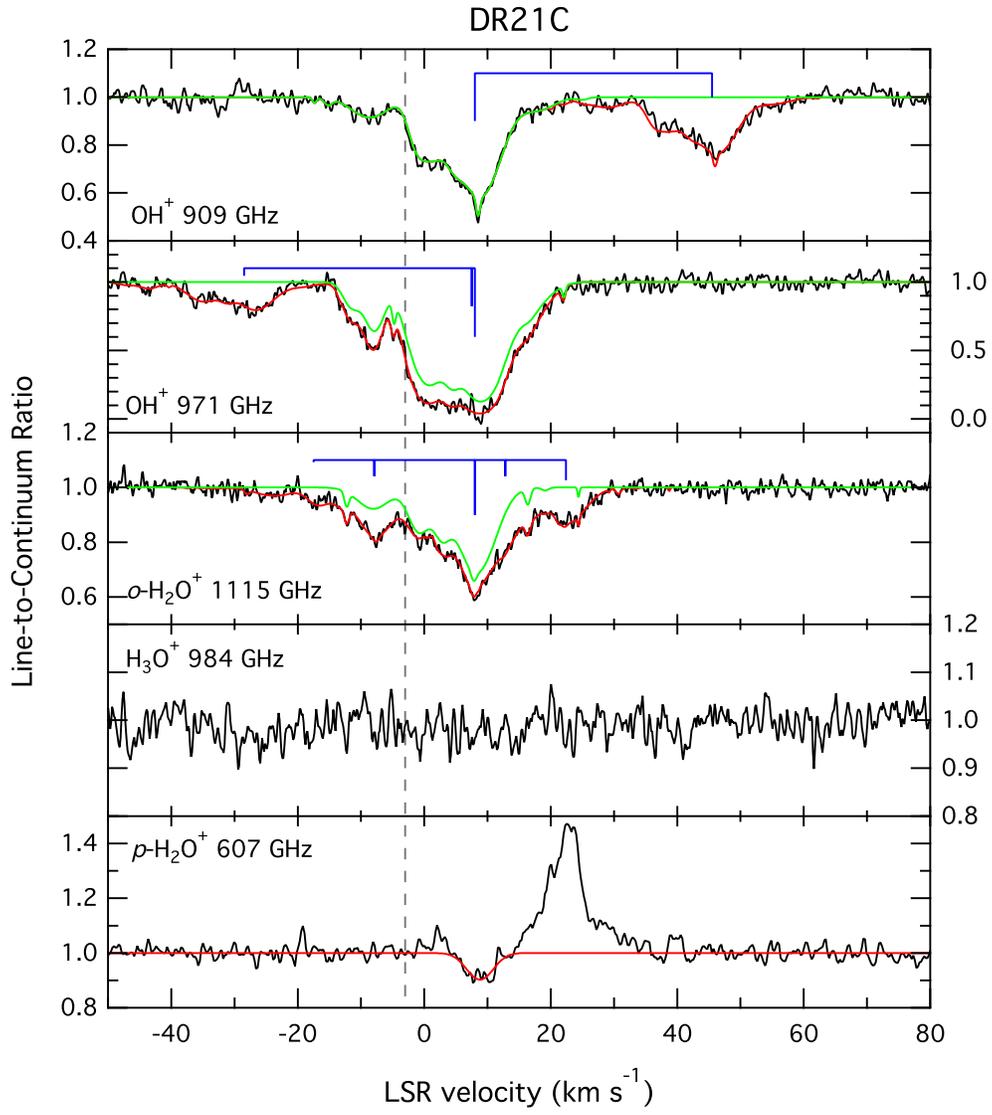}
\caption{Same as Figure \ref{fig_sgra20_spectra} but for DR21C.}
\label{fig_dr21_spectra}
\end{figure}

\begin{figure}
\epsscale{0.8}
\plotone{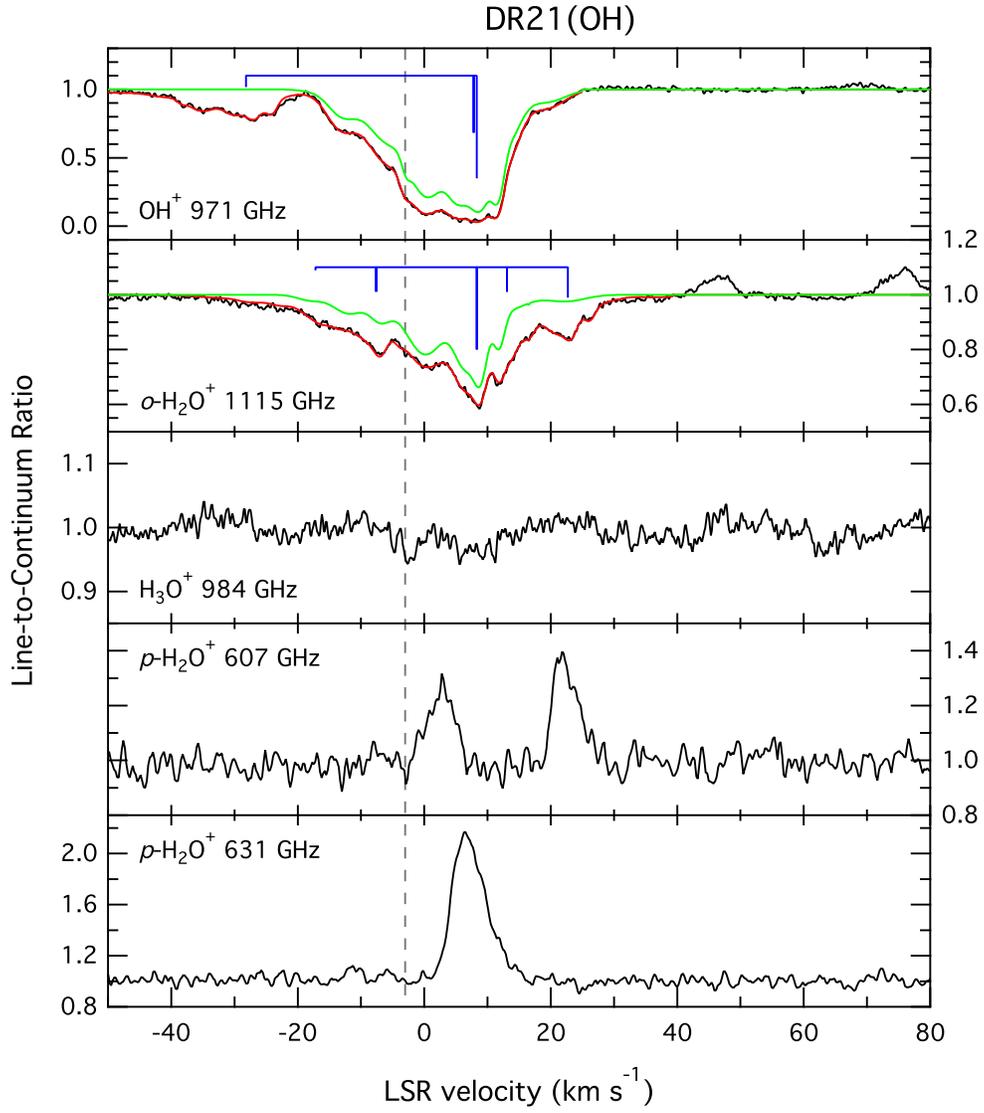}
\caption{Same as Figure \ref{fig_sgra20_spectra} but for DR21(OH).}
\label{fig_dr21oh_spectra}
\end{figure}

\begin{figure}
\epsscale{0.8}
\plotone{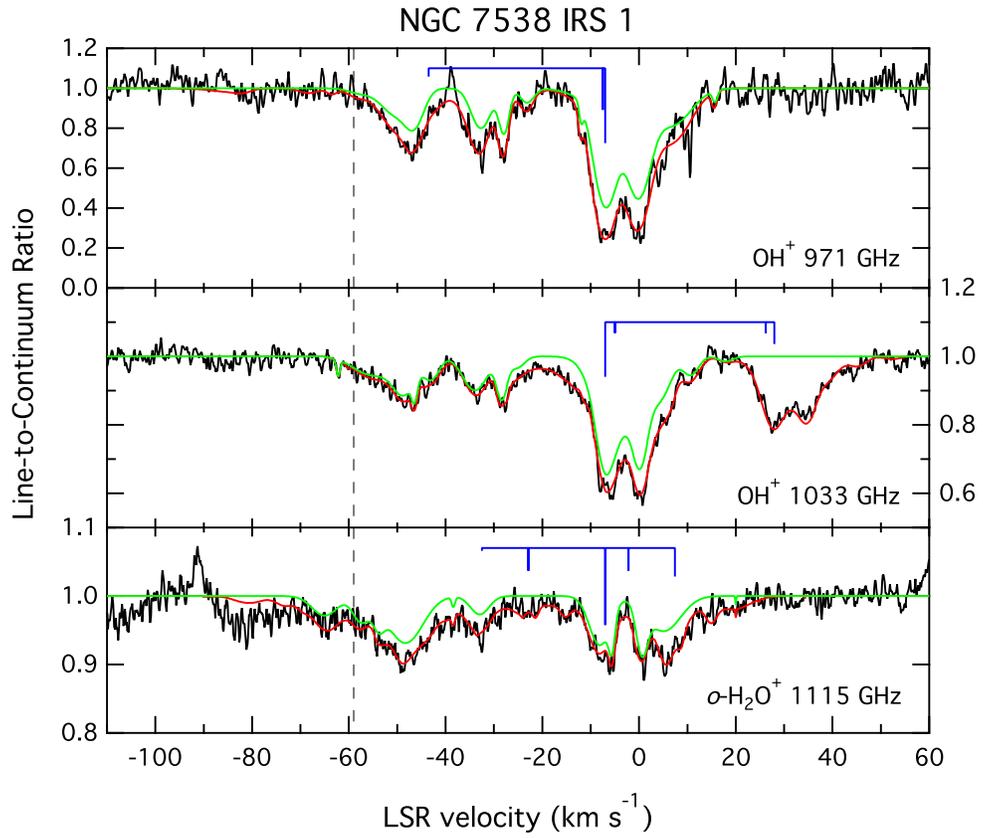}
\caption{Same as Figure \ref{fig_sgra20_spectra} but for NGC 7538 IRS 1.}
\label{fig_ngc7538_spectra}
\end{figure}

\begin{figure}
\epsscale{0.8}
\plotone{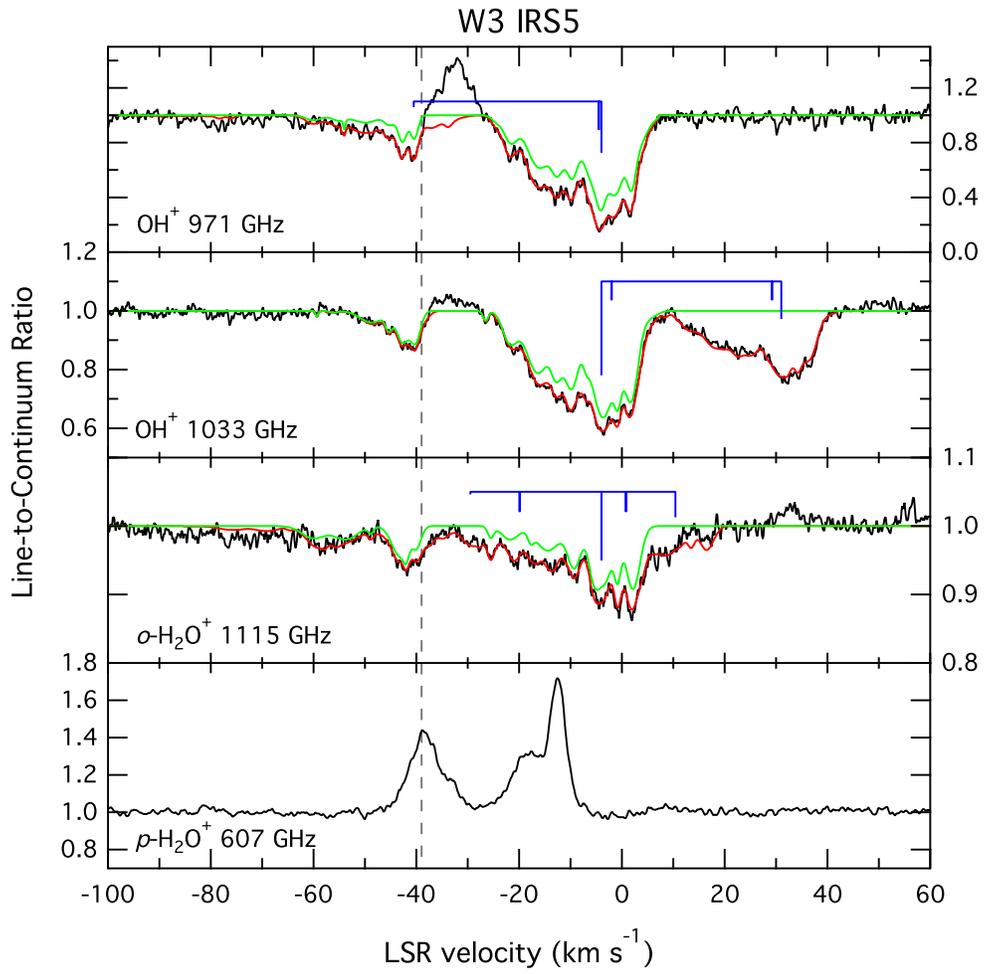}
\caption{Same as Figure \ref{fig_sgra20_spectra} but for W3 IRS5.}

\label{fig_w3irs5_spectra}
\end{figure}

\begin{figure}
\epsscale{0.8}
\plotone{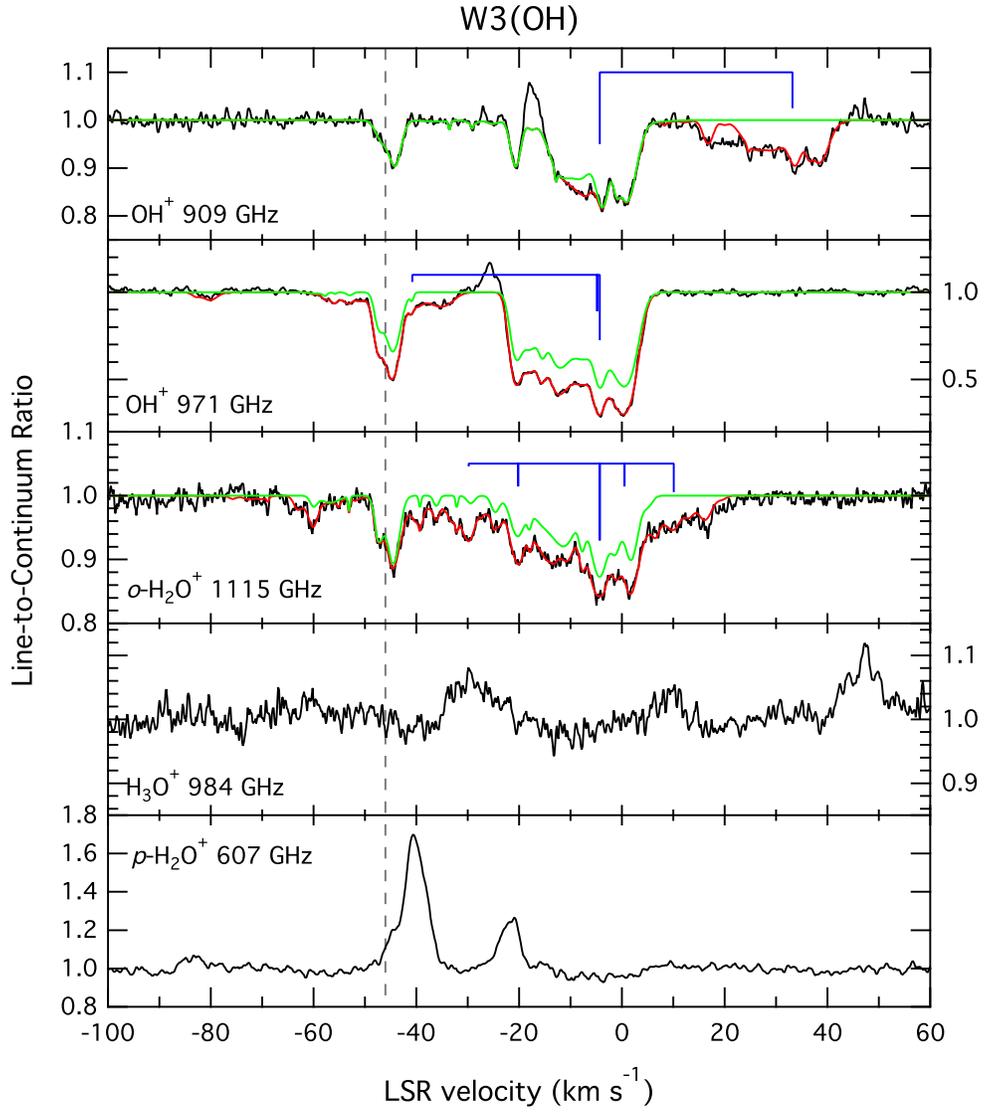}
\caption{Same as Figure \ref{fig_sgra20_spectra} but for W3(OH).}
\label{fig_w3oh_spectra}
\end{figure}

\begin{figure}
\epsscale{0.8}
\plotone{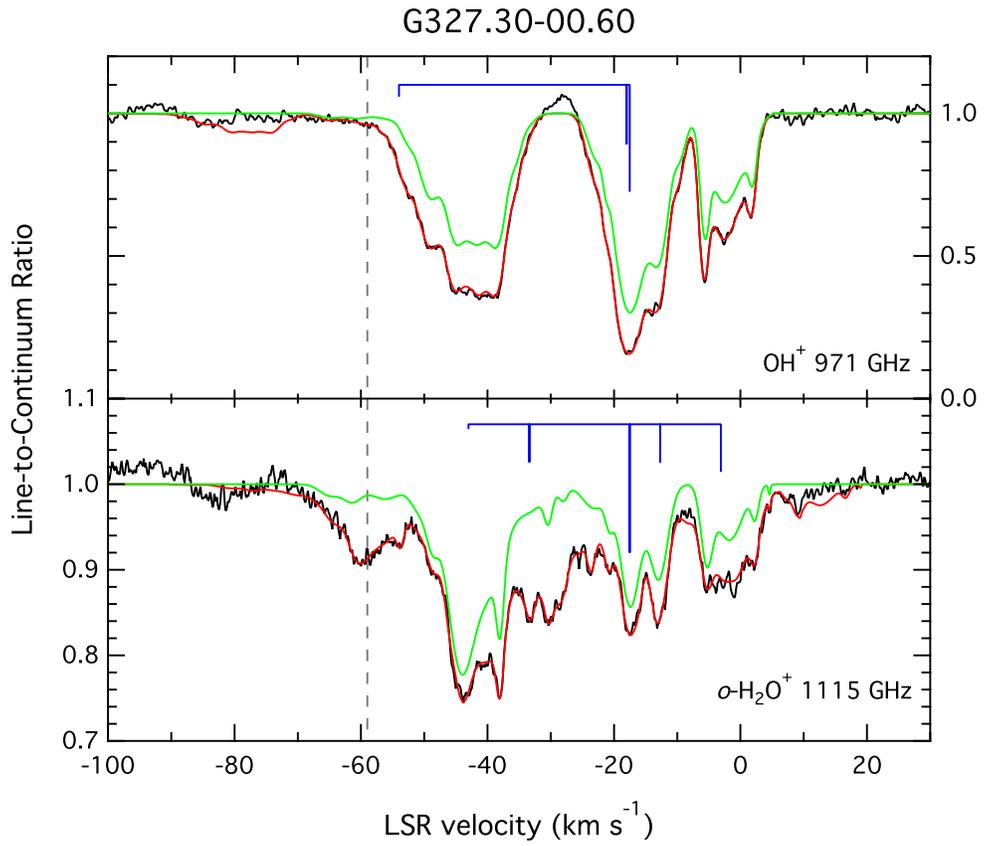}
\caption{Same as Figure \ref{fig_sgra20_spectra} but for G327.30$-$00.60.}
\label{fig_g327_spectra}
\end{figure}

\begin{figure}
\epsscale{0.8}
\plotone{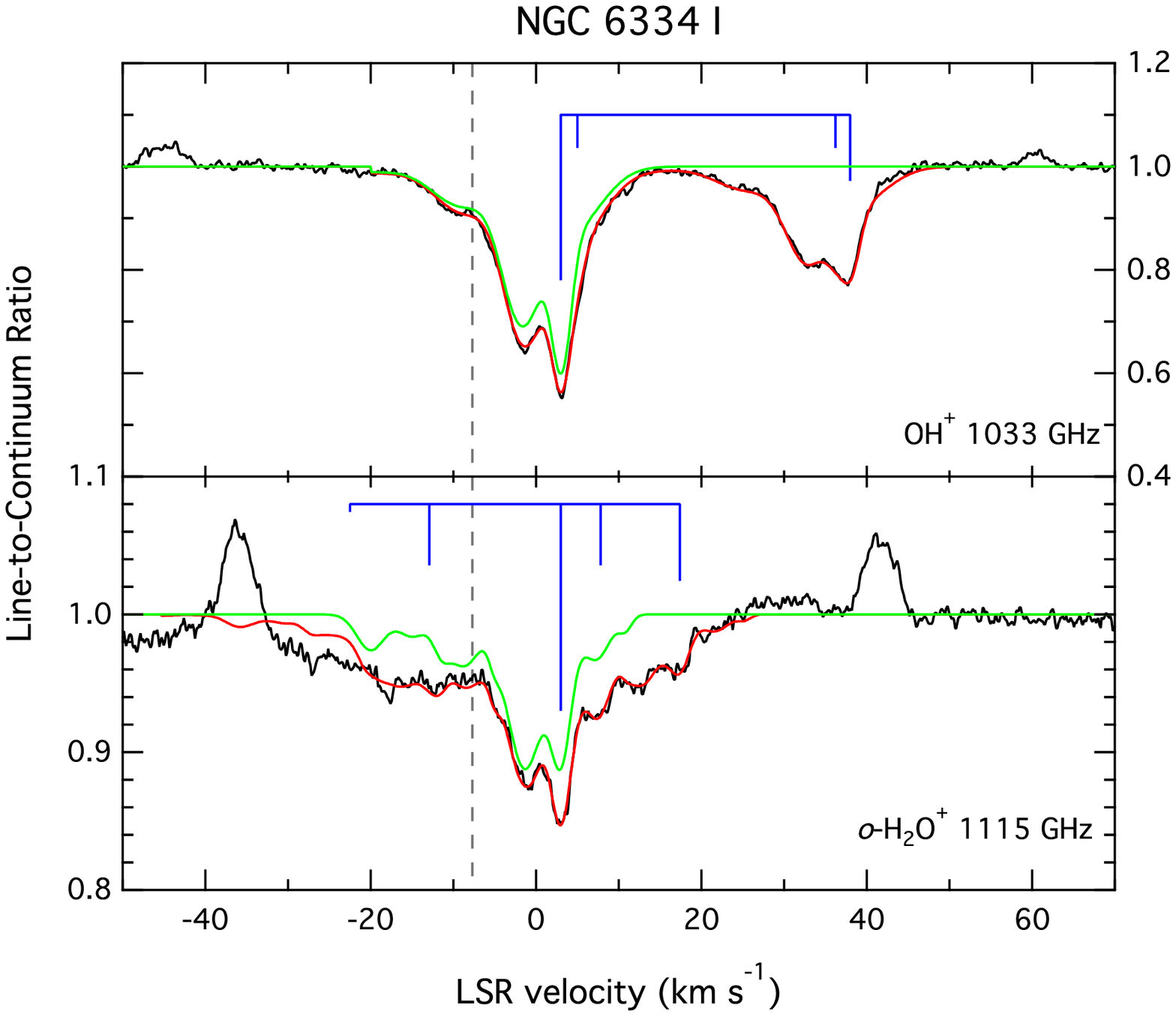}
\caption{Same as Figure \ref{fig_sgra20_spectra} but for NGC 6334 I.}
\label{fig_ngc6334I_spectra}
\end{figure}

\begin{figure}
\epsscale{0.8}
\plotone{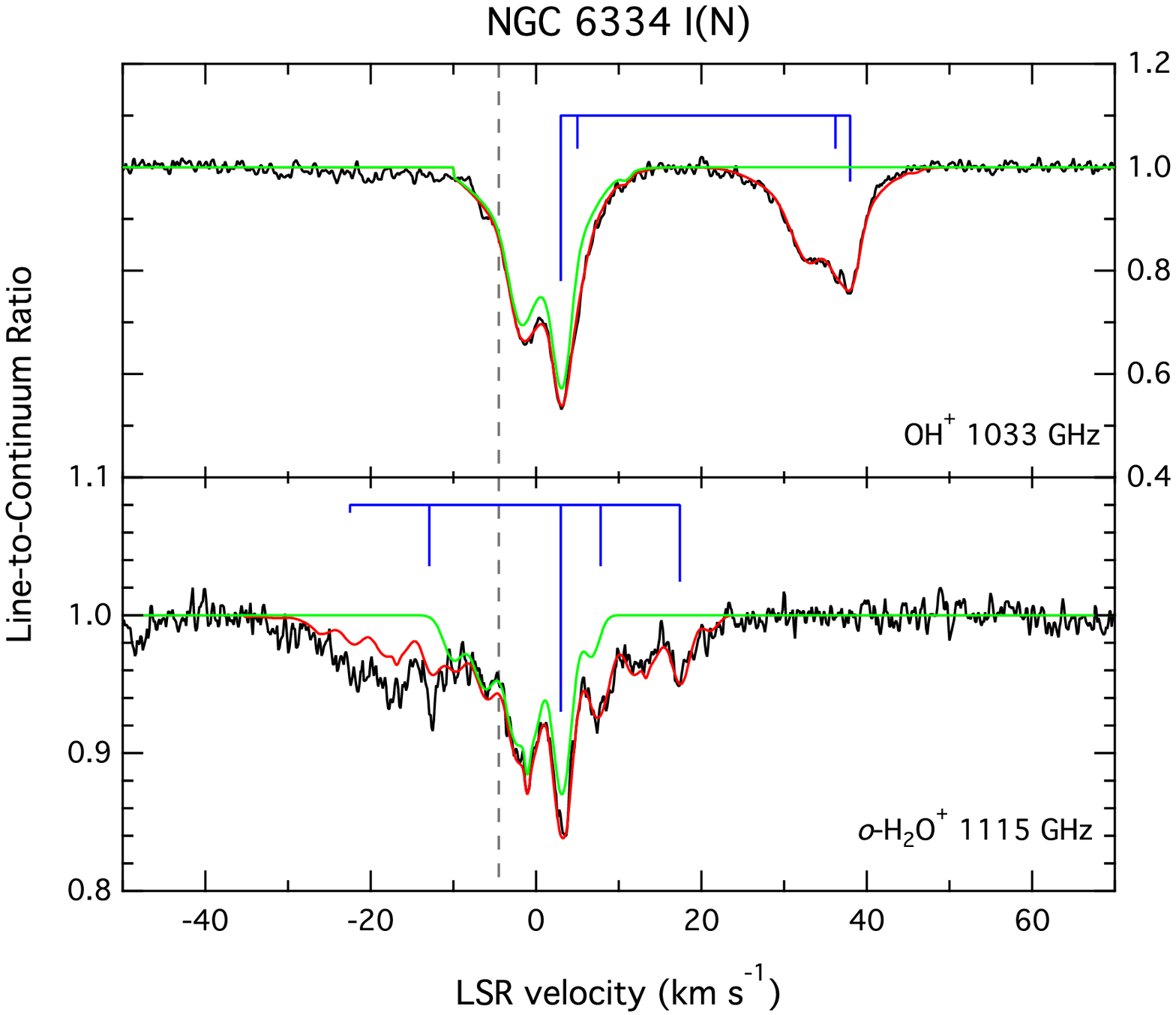}
\caption{Same as Figure \ref{fig_sgra20_spectra} but for NGC 6334 I(N).}
\label{fig_ngc6334In_spectra}
\end{figure}

The basic fitting procedure used in our analysis has been described previously by \citet{neufeld2010}, but due to some differences we briefly review it here.  Absorption features are assumed to result from the combination of multiple components with Gaussian opacity profiles.  Each component is defined by a centroid velocity, velocity full-width at half-maximum, and maximum optical depth which act as variables in the fitting process. For transitions with hyperfine structure each component consists of multiple Gaussians in opacity.  The strongest hyperfine feature is defined as above, and the other hyperfine features are forced to have the same velocity width, with fixed relative intensities and fixed velocity separations---with respect to the strongest feature---defined by transition frequencies, statistical weights, and spontaneous emission coefficients.\footnote{Imagine convolving the hyperfine structure stick diagrams in Figures \ref{fig_sgra20_spectra}--\ref{fig_ngc6334In_spectra} with a Gaussian line profile to picture absorption from a single velocity component.}  Some number of velocity components (between 2 and 20 depending on the complexity of the absorption profile) is initially chosen, and the sum of those components is used to fit the absorption profile.  The number of components is then revised as needed to produce a reasonable fit to the spectra.  These fits are shown as the red curves in Figures \ref{fig_sgra20_spectra}--\ref{fig_sgra50_spectra} and \ref{fig_w28a_spectra}--\ref{fig_ngc6334In_spectra}. To determine the actual distribution of molecules in velocity space when considering a transition with hyperfine splitting, we examine the portion of the fit caused only by the strongest hyperfine feature, as shown by the green curves in Figures \ref{fig_sgra20_spectra}--\ref{fig_ngc6334In_spectra}).
  
\subsection{Column Density}

From the above fitting procedure we determine the optical depth and differential column density ($dN/dv$) as functions of LSR velocity along a line of sight.  The column density in any velocity range can then be determined by integrating $dN/dv$ over that range.  Using the OH$^+$ and $o$-H$_2$O$^+$ absorption profiles (green curves) we select velocity intervals that correspond to what appear to be separate absorption components, and integrate $dN/dv$ over those intervals.  
Column densities determined from this analysis for all species are reported in Tables \ref{tbl_columns} and \ref{tbl_results}. In all cloud components we assume nearly all molecules are in the ground rotational state for the purpose of determining the total column density of OH$^+$ and H$_2$O$^+$ from our observations. Gas densities in diffuse clouds are sufficiently low that collisional excitation is unimportant, and spontaneous radiative decay rates for the studied transitions are large (see Table \ref{tbl_transitions}), so the excitation temperature is very likely controlled by the cosmic microwave background radiation (i.e., $T_{ex}\sim2.7$~K).  The assumption that nearly all OH$^+$ and H$_2$O$^+$ molecules are in the ground rotational state is thus justified in the diffuse ISM, although in components where the molecules reside in gas that is part of the envelope surrounding the H~\textsc{ii} regions used as background sources this may no longer be the case.\footnote{These velocity intervals are identified in Tables \ref{tbl_columns} and \ref{tbl_results}.} For H$_3$O$^+$, most of our observations do not probe the lowest lying state, nor is the above assumption valid, so we only report state-specific column densities.  In cases where multiple transitions of a given species are observed, the column densities determined from individual transitions are weighted by $1/\sigma^2$ (i.e., inverse of the square of the standard deviation presented as uncertainty in Table \ref{tbl_columns}) when determining the average column density for that species (e.g., OH$^+$ toward G029.96$-$00.02).  If multiple transitions of a species are observed and one transition is saturated (e.g., OH$^+$ toward W31C and W49N), only the unsaturated transition is used to determine the column density.  When $p$-H$_2$O$^+$ is not detected, an {\it ortho}-to-{\it para} ratio (OPR) of 3 is assumed in determining $N({\rm H_{2}O^+})$ from $N(o$-${\rm H_{2}O^+})$. Only for the Sgr~B2 sight lines is a different analysis used, where absorption features caused by all transitions of a given species (e.g., 909~GHz, 971~GHz, and 1033~GHz for OH$^+$; 1115~GHz and 1139~GHz for $o$-H$_2$O$^+$) are fit simultaneously to determine $dN/dv$ \citep{schilke2013}.

\subsection{Molecular Hydrogen Fraction} \label{subsec_fH2}

A steady-state analysis of the H$_2$O$^+$ abundance governed by reactions (\ref{reac_OH+_H2}), (\ref{reac_H2O+_H2}), and (\ref{reac_H2O+_e}) gives the equation
\begin{equation}
n({\rm OH}^+)n({\rm H}_2)k_4=n({\rm H_2O}^+)[n({\rm H}_2)k_6+n(e)k_7],
\label{eq_H2O+_steadystate}
\end{equation}
where $k_i$ is the rate coefficient\footnote{Rate coefficients are taken from the UMIST database for astrochemistry \citep{mcelroy2013}.} for reaction $i$ within this paper.  Through substitutions and rearrangement as shown in \citet{indriolo2012w51} this is solved for the molecular hydrogen fraction,
\begin{equation}
f_{\rm H_{2}}=\frac{2x_{e}k_7/k_4}{N({\rm OH^{+}})/N({\rm H_{2}O^{+}})-k_6/k_4}.
\label{eq_hydride_fH2}
\end{equation} 
This assumes constant densities (for the conversion from number density to column density) and constant temperature ($k_7$ is temperature dependent) over the region probed.  In determining $f_{\rm H_2}$ we take $x_e=1.5\times10^{-4}$ \citep[assuming $x_e=x({\rm C}^+)$;][]{cardelli1996,sofia2004} and $T=100$~K (typical value of the H~\textsc{i} spin temperature) in computing the various reaction rate coefficients.  Resulting values are presented in Table \ref{tbl_results}.


\subsection{Cosmic-Ray Ionization Rate} \label{subsec_ionizationrate}

A similar analysis of steady-state chemistry for OH$^+$ gives
\begin{equation}
\epsilon\zeta_{\rm H}n({\rm H})=n({\rm OH}^+)[n({\rm H}_2)k_4+n(e)k_5].
\label{eq_OH+_steadystate}
\end{equation}
This equation accounts for the dominant reaction partners by which OH$^+$ is destroyed (i.e., H$_2$ and electrons), but not every instance of hydrogen ionization results in the formation of OH$^+$ due to the backward version of reaction (\ref{reac_Hp_Op}) and the neutralization of protons on dust grains and PAHs \citep{wolfire2003,liszt2003,hollenbach2012}.  To accommodate this fact, we follow \citet{neufeld2010} in introducing the efficiency factor, $\epsilon$, on the left-hand side of equation (\ref{eq_OH+_steadystate}).  Given the same assumptions as above, substitution and rearrangement leads to 
\begin{equation}
\epsilon\zeta_{\rm H}=\frac{N({\rm OH}^+)}{N({\rm H})}n_{\rm H}\left[\frac{f_{\rm H_2}}{2}k_4+x_ek_5 \right],
\label{eq_OH+_zeta}
\end{equation}
and the substitution of equation (\ref{eq_hydride_fH2}) for $f_{\rm H_2}$ to
\begin{equation}
\epsilon\zeta_{\rm H}=\frac{N({\rm OH}^+)}{N({\rm H})}n_{\rm H}x_e\left[\frac{k_7}{N({\rm OH^{+}})/N({\rm H_{2}O^{+}})-k_6/k_4}+k_5 \right].
\label{eq_zeta_n}
\end{equation}
In addition to the column densities of OH$^+$ and H$_2$O$^+$, this analysis also requires the column density of atomic hydrogen, $N({\rm H})$, and values are reported in Table \ref{tbl_results}.  In several of our target sight lines $N({\rm H})$ is determined from 21~cm absorption observations analyzed by Winkel et al. 2015 (in prep), and in the remainder we have used H~\textsc{i} spectra reported in various works (see Table \ref{tbl_results} description for references).  The number density of hydrogen nuclei is also needed in calculating the ionization rate, and we adopt $n_{\rm H}=35$~cm$^{-3}$ following the reasoning in \citet{indriolo2012w51}.  This value arises from assuming that the diffuse atomic outer layers of a cloud with $T=100$~K are in pressure balance with the diffuse molecular interior with $T=70$~K and $n_{\rm H}=100$~cm$^{-3}$, and it is in very good agreement with the mean thermal pressure ($\log(P/k)=3.58)$ inferred from fine-structure excitation of C~\textsc{i} in diffuse clouds \citep{jenkins2011}.  Still, the determination of interstellar densities is highly uncertain, and we discuss the effects this may have on our analysis below.

Observed column densities of OH$^+$, H$_2$O$^+$, and H are then used in concert with rate coefficients and adopted values of $T$, $x_e$, and $n_{\rm H}$ to calculate $\epsilon\zeta_{\rm H}$ in each cloud component.  To convert $\epsilon\zeta_{\rm H}$ to the cosmic-ray ionization rate the efficiency factor must be known, and we adopt $\epsilon=0.07$ as found during our previous study of the W51e sight line where H$_3^+$ observations were used to independently determine the ionization rate and calibrate $\epsilon$ \citep{indriolo2012w51}.  The value of $\epsilon=0.07\pm0.04$ presented in \citet{indriolo2012w51} is the only observational determination of the efficiency factor, although it has also been computed as part of chemical models studying OH$^+$, H$_2$O$^+$, and H$_3$O$^+$ presented by \citet{hollenbach2012}.  Those authors find $0.05\lesssim \epsilon\lesssim 0.2$, with the value changing for different densities, ionization rates, depth into a cloud, etc.  While our single observational determination of $\epsilon$ falls within the range based on chemical modeling, there is clearly still large uncertainty, and we discuss below the effects that variations in $\epsilon$ have on our analysis.  Assuming $\epsilon=0.07$ we calculate the cosmic-ray ionization rate of atomic hydrogen, and values of $\zeta_{\rm H}$ are presented in Table \ref{tbl_results}. 

\subsection{Uncertainties in $T$, $x_e$, $n_{\rm H}$, and $\epsilon$} \label{subsec_unc}
During our analysis we have made various assumptions regarding certain variables, namely that $x_e=1.5\times10^{-4}$, $n_{\rm H}=35$~cm$^{-3}$, $T=100$~K, and $\epsilon=0.07$ in all cloud components.  Here we discuss the uncertainties associated with these parameters.

Temperature (specifically gas kinetic temperature) can affect the rates at which certain chemical reactions occur.  The hydrogen abstraction reactions relevant to our analysis (\ref{reac_Op_H2}, \ref{reac_OH+_H2}, and \ref{reac_H2O+_H2}) are temperature independent, while dissociative recombination with electrons (reactions \ref{reac_OH+_e} and \ref{reac_H2O+_e}) is only weakly dependent on temperature ($k\propto T^{-0.5}$).  Inferred temperatures for diffuse molecular clouds do not vary widely ($T\approx60$--120~K), and we are observing species primarily in the warmer outer regions of these clouds, so uncertainties in $T$ should not significantly affect our results.

The electron fraction is frequently approximated by the C$^+$ fractional abundance in diffuse clouds where singly ionized carbon is responsible for the majority of free electrons.  This has been found to be about $1.5\times10^{-4}$ with moderate variance across observed sight lines that probe gas within about 1~kpc of the Sun \citep{cardelli1996,sofia2004}.  Metallicities tend to increase at smaller Galactocentric radii \citep{rolleston2000}, and a typical assumed gradient in carbon abundance results in a factor of 2--3 increase in $x({\rm C})$ at $R_{\rm gal}=3$~kpc \citep[e.g.,][]{wolfire2003,pineda2013,langer2014}.  It is not clear how exactly $x_e$ changes with Galactocentric radius, but it is reasonable to assume variations of about a factor of 3 with respect to the adopted value across our sample due to variations in the carbon abundance.


While the contribution to the electron abundance from ionized species other than C$^+$ (e.g., Si$^+$, S$^+$) is generally negligible in diffuse gas, as $\zeta_{\rm H}$ increases the H$^+$ abundance can become comparable to or exceed that of C$^+$.  A prescription for calculating the steady-state value of $x_e$ as a function of $\zeta_{\rm H}$ in purely atomic gas is given by \citet{draine2011}, and includes the effects of grain-assisted recombination \citep{weingartner2001}.  The resulting electron fraction is dependent on input parameters such as temperature, density, and interstellar radiation field, and for our assumed density $x_e$ is relatively constant for $\zeta_{\rm H}\lesssim 10^{-16}$~s$^{-1}$, and increases roughly as $\sqrt{\zeta_{\rm H}}$ for larger ionization rates.  In this regime, the approximation in equations (\ref{eq_OH+_zeta}) and (\ref{eq_zeta_n}) that $x_e$ is independent of $\zeta_{\rm H}$ begins to break down, and $N({\rm OH}^+)$ no longer increases linearly with $\zeta_{\rm H}$ \citep[similar to findings for H$_3^+$;][]{liszt2007,goto2008}.  In components where we find high ionization rates, $x_e$ may in fact be significantly larger than we have assumed (up to a factor of about 20 for $\zeta_{\rm H}\sim10^{-14}$~s$^{-1}$), which would in turn give even larger ionization rates and larger molecular hydrogen fractions than we have reported.  Still, given the relationship between $x_e$ and $\zeta_{\rm H}$, the values we report remain valid lower limits.

Interstellar gas densities are difficult to constrain and typically have large uncertainties.  Estimates of $n_{\rm H}$ are often made using the relative populations in excited states of atoms and molecules.  A recent analysis of CO observations gives densities in the range $n_{\rm H}\approx 20$--200~cm$^{-3}$ \citep{goldsmith2013}, while C$_2$ observations result in $n_{\rm H}\approx 100$--400~cm$^{-3}$ \citep{sonnentrucker2007} for diffuse molecular clouds. Excitation of the fine-structure levels of O~\textsc{i} has been used to find $n_{\rm H}\approx 5$--25~cm$^{-3}$ in diffuse gas \citep{sonnentrucker2002,sonnentrucker2003}, while excitation of C~\textsc{i} has been used to determine thermal pressures that are consistent with $n_{\rm H}\approx 10$--100~cm$^{-3}$ \citep{jenkins2001,jenkins2011}.  Observations of C$^{+}$ in many of the same sight lines studied herein are used to infer $n_{\rm H}\approx 40$--100~cm$^{-3}$ \citep{gerin2014}.   Clearly, there is large variance among inferred densities for diffuse clouds, although much of this may be due to the specific region probed (i.e., molecular interior versus atomic exterior).  If we take the extreme values given above as limits, our adopted value of $n_{\rm H}\approx 35$~cm$^{-3}$ can vary up or down by roughly a factor of 10 between different components.  However, given the density estimates from \citet{gerin2014} and pressures inferred by \citet{jenkins2011}, it seems likely that this uncertainty is more commonly only a factor of 3 or so in the diffuse foreground gas along our targeted sight lines.\footnote{Note that we do not simply adopt densities determined from C$^+$ as this species also traces gas with large $f_{\rm H_2}$ where densities are likely higher than the regions containing most of the OH$^+$ and H$_2$O$^+$.}

The efficiency factor $\epsilon$ (fraction of instances where cosmic-ray ionization of H leads to formation of OH$^+$) has only a single constraint via observations ($\epsilon=0.07\pm0.04$), and a few estimates from chemical models focused on oxygen-bearing species ($0.05\leq\epsilon\leq0.2)$.  This parameter is affected by gas conditions (e.g., $T$, $n_{\rm H}$, relative abundances), and will vary between different clouds.  Our best estimate on the uncertainty in $\epsilon$ comes from the chemical models of \citet{hollenbach2012}, so we consider limits of about a factor of 2 below and a factor of 3 above the adopted value of $\epsilon=0.07$.



\subsection{Kinematic Distances} \label{subsec_kind}

In order to explore any correlation that $f_{\rm H_2}$ or $\zeta_{\rm H}$ may have with Galactocentric radius we use a kinematic analysis to estimate $R_{gal}$ for the various velocity intervals of absorbing gas along each line of sight.  We adopt the functional form of the rotation curve presented by \citet{persic1996}, and use the input parameters recommended by \citet{reid2014}, including a distance to the Galactic center of 8.34~kpc and a rotation speed of 240~km~s$^{-1}$. This allows us to determine the expected line-of-sight velocity as a function of $R_{gal}$ given the Galactic longitude of each sight line. Within each velocity interval along a sight line we choose a single velocity---usually corresponding to maximum absorption---which is used in determining $R_{gal}$ for that component.  From $R_{gal}$ we determine the near and far kinematic distances to each absorption component.  In several cases, departures from the expected velocity due to peculiar motions cause this analysis to produce unphysical results.  When the resulting Galactocentric radius is smaller than the radius of the tangent point along a line of sight, we set $R_{gal}$ equal to the tangent point radius.  When the kinematic distance is larger than the assumed distance of the background source (see Table \ref{tbl_target} and Section \ref{subsec_los}) we set the distance equal to that of the background source and correct $R_{gal}$ accordingly.  When the kinematic distance is smaller than 0.1~kpc we set it equal to 0.1~kpc (and adjust $R_{gal}$) so that the gas is outside of the local bubble.  For most sight lines the kinematic distance ambiguity is solved because the background source is on the near side of the tangent point.  In cases where the ambiguity remains (e.g., W49N) both distances are reported.  Some of our target sight lines have distance estimates to foreground clouds available in the literature, which we take to be more robust than our own kinematic analysis.  For these sight lines (M$-$0.13$-$0.08, M$-$0.02$-$0.07, Sgr~B2(M), Sgr~B2(N), W31C, W3~IRS5, and W3(OH); see sections describing individual sight lines for references) we adopt previously determined distances.  In the case of AFGL~2591, we assume both absorption components are associated with the background source itself.  Galactocentric radii and distances for each velocity component are reported in Table \ref{tbl_results}.

\section{DISCUSSION} \label{sec_disc}

As described above, spectra for each sight line were divided into velocity intervals roughly corresponding to absorption features for the purpose of analyzing our data.  A total of 105 separate components containing OH$^+$ absorption are defined in our sample, of which 100 also show $o$-H$_2$O$^+$ absorption.  In contrast, H$_3$O$^+$ absorption is only seen in 16 components (12 of which are in the Galactic center), and $p$-H$_2$O$^+$ absorption in 11 components (4 of which are in the Galactic center). In many cases though,
potential for detection of the $p$-H$_2$O$^+$ 607~GHz line is impeded by emission from the $J=7$--6 transition of H$^{13}$CO$^+$ (607.1747~GHz) and the $J_{K_{a},K_{c}}=12_{2,10}$--11$_{1,10}$ transition of CH$_3$OH (607.2158~GHz).  Similarly, the 631~GHz transition of $p$-H$_2$O$^+$ is often obscured by emission from the $J_{K_{a},K_{c}}=9_{1,9}$--8$_{1,8}$ transition of H$_2$CO (631.7028~GHz).  The rest frequency of the H$_3$O$^+$ 1655~GHz transition is only 34~MHz below (6~km~s$^{-1}$ redshift) that of the 2$_{1,2}$--1$_{0,1}$ transition of H$_2^{18}$O, making a confirmed detection difficult.  Only in two sight lines (W31C and W49N) do we present the 1655~GHz spectra, as in all others where the transition was covered we are confident the absorption signals are due to H$_2^{18}$O \citep[identification aided by absorption profiles of the 1$_{1,1}$--0$_{0,0}$ transition of H$_2^{18}$O presented in][]{vandertak2013H2O}. A more detailed description of our findings in each line of sight follows.

\subsection{Line of Sight Properties} \label{subsec_los}

\subsubsection{M$-$0.13$-$0.08 and M$-$0.02$-$0.07}

Two well known molecular clouds in the Galactic center region are M$-$0.13$-$0.08 and M$-$0.02$-$0.07, also commonly referred to as the Sgr A +20~km~s$^{-1}$ and Sgr A +50~km~s$^{-1}$ clouds due to their respective radial velocities.  Both are within 10~pc of the Galactic center \citep{ferriere2012}, which is 8.34$\pm$0.16~kpc away from the Sun \citep{reid2014}, and are separated by a projected distance of 15.6~pc.  Each cloud has associated compact H~\textsc{ii} regions---source SgrA--G from \citet{ho1985} in the case of M$-$0.13$-$0.08, and SgrA--A, SgrA--B, SgrA--C, and SgrA--D (also referred to as the G$-$0.02$-$0.07 complex) in the case of M$-$0.02$-$0.07 \citep{ekers1983,mills2011}---but it is cool dust that serves as the background continuum sources for these sight lines.

M$-$0.13$-$0.08 shows OH$^+$ absorption across the entire velocity range from $-$210~km~s$^{-1}$ to 30~km~s$^{-1}$ (Figure \ref{fig_sgra20_spectra}).  Absorption at $v_{\rm LSR}\lesssim-60$~km~s$^{-1}$ is thought to be due entirely to gas in the central molecular zone (CMZ) within the Galactic center region \citep{sonnentrucker2013}, while at $v_{\rm LSR}\gtrsim-60$~km~s$^{-1}$ there is some combination of foreground spiral arms that absorb at distinct velocities and gas in the CMZ.  We attribute absorption at $-50$~km~s$^{-1} \leq v_{\rm LSR}\leq -40$~km~s$^{-1}$  to the 3~kpc spiral arm \citep{dame2008}, and at $-40$~km~s$^{-1} \leq v_{\rm LSR}\leq -15$~km~s$^{-1}$ to the 4.5~kpc spiral arm \citep{menon1970}, although both intervals are likely contaminated by gas in the Galactic center as well.  Absorption from $-$15~km~s$^{-1}$ to 30~km~s$^{-1}$ is due to some combination of local gas and the CMZ, including the cloud in which the continuum source is embedded (i.e., the +20~km~s$^{-1}$ cloud). The $o$-H$_2$O$^+$ spectrum follows a similar pattern, but with weaker absorption in many components.  H$_3$O$^+$ absorption is only seen in a narrow component at 12~km~s$^{-1}$, coming from the molecular cloud itself.\footnote{The spectrum is truncated below $v_{\rm LSR}\leq-90$~km~s$^{-1}$ due to interference from the 971~GHz transition of OH$^+$ in the other sideband.}

Similarly, M$-$0.02$-$0.07 shows OH$^+$ and $o$-H$_2$O$^+$ absorption from $-$210~km~s$^{-1}$ to 70~km~s$^{-1}$ (Figure \ref{fig_sgra50_spectra}).  We assume roughly the same breakdown between CMZ and foreground gas, with the 3~kpc spiral arm at $-61$~km~s$^{-1} \leq v_{\rm LSR}\leq -47$~km~s$^{-1}$, the 4.5~kpc spiral arm at $-47$~km~s$^{-1} \leq v_{\rm LSR}\leq -13$~km~s$^{-1}$, and the CMZ at $v_{\rm LSR}\leq-61$~km~s$^{-1}$ and $v_{\rm LSR}\geq-13$~km~s$^{-1}$. The background molecular cloud (Sgr A +50~km~s$^{-1}$) is responsible for absorption between 20~km~s$^{-1}$ and 70~km~s$^{-1}$.  Again, the background source is the only component that shows substantial H$_3$O$^+$ absorption, although there may also be a weak feature at $-140$~km~s$^{-1}$. Both the 607~GHz and 631~GHz transitions of $p$-H$_2$O$^+$ were also targeted toward M$-$0.02$-$0.07, and, despite high noise levels, we consider the features at $-140$~km~s$^{-1}$ that coincide with the strongest $o$-H$_2$O$^+$ absorption to be detections.  Values of $N(p{\rm -H_2O^+})$ derived from both transitions are in agreement.

It must be noted that for both sight lines blending of absorption from foreground spiral arms and from the CMZ complicates our analysis.  We have attributed absorption in select velocity intervals entirely to foreground clouds following previous studies of molecular absorption toward the Galactic center \citep[e.g.,][]{monje2011,sonnentrucker2013,schilke2010,schilke2014}, but other studies have shown that the entire velocity range under consideration also contains absorption from gas in the CMZ \citep[e.g.,][]{oka2005,geballe2010,goto2011,goto2014}.  Results inferred from absorption in these velocity ranges---i.e., those assigned to the 3~kpc and 4.5~kpc spiral arms---should be viewed with caution. The same is true for select velocity intervals in the Sgr~B2 sight lines discussed below.

\subsubsection{Sgr B2(M) and Sgr B2(N)}

Sgr~B2 is a giant molecular cloud within the Galactic center region that contains multiple cores---including Sgr B2(M) and Sgr B2(N)---where prolific star formation is occurring.  Different studies place Sgr~B2 in front of \citep{reid2009_sgrb2} or behind \citep{molinari2011} Sgr~A*, but always within $\sim150$~pc, and we adopt $d=8.34$~kpc, as the precise location is not vital to our study.  At this distance the projected separation between Sgr~B2 and Sgr~A* is about 100~pc, and the projected separation between the Sgr~B2(M) and Sgr~B2(N) cores is 1.8~pc.

Spectra of OH$^+$, H$_2$O$^+$, and H$_3$O$^+$ toward Sgr~B2(M) and Sgr~B2(N) are largely similar (Figures \ref{fig_sgrb2m_spectra} and \ref{fig_sgrb2n_spectra}) with strong absorption extending from about $-120$~km~s$^{-1}$ to 40~km~s$^{-1}$.  Absorption across this entire velocity range is likely caused by gas within the Galactic center and foreground spiral arms that contribute at specific velocities.  The lack of sharp, well-defined features makes it difficult to attribute absorption to any particular spiral arm, but we assume that absorption in the $-60$~km~s$^{-1}\lesssim v_{\rm LSR}\lesssim-30$~km~s$^{-1}$ interval arises in the 3~kpc arm, and in the $-30$~km~s$^{-1}\lesssim v_{\rm LSR}\lesssim-5$~km~s$^{-1}$ interval in the 4.5~kpc arm, with the caveat that there is likely considerable contamination from gas within the Galactic center as well. Systemic velocities of the background sources differ slightly, about 63~km~s$^{-1}$ for Sgr~B2(M) and 66~km~s$^{-1}$ for Sgr~B2(N), and both sources show absorption, but Sgr~B2(N) has an additional absorption component near 80~km~s$^{-1}$ seen only in H$_3$O$^+$.  The Sgr~B2 sight lines are unique in our survey in that H$_3$O$^+$ absorption is detected in all velocity components. While we only list column densities in the 1$_0^+$ state, a much more thorough analysis utilizing transitions out of 11 levels of H$_3$O$^+$ in these sight lines (beyond the scope of this paper) has been carried out by \citet{lis2014}. Our reported values for $N(1_0^+)$ are in good agreement with theirs.

Although several spectra of both {\it ortho} and {\it para} H$_2$O$^+$ toward Sgr~B2(M) have been presented and analyzed in previous studies \citep{ossenkopf2010,schilke2010,schilke2013}, we reproduce the 1115~GHz and 1139~GHz absorption lines here to facilitate comparison with OH$^+$. Both the 971~GHz and 1033~GHz lines of OH$^+$ are saturated, and knowledge of the velocity structure is almost entirely dependent on the 909~GHz transition. Still, the OH$^+$ and $o$-H$_2$O$^+$ profiles are nearly identical in velocity structure for both Sgr~B2(M) and Sgr~B2(N). These sight lines are also unique in that $p$-H$_2$O$^+$ is detected in all velocity components as shown by \citet{schilke2013}, and where available we use column densities determined from that study in computing the OPR shown in Table \ref{tbl_columns}, as well as total $N({\rm H_2O}^+)$.

\subsubsection{W28A}
The ultracompact H~\textsc{ii} region W28A (also known as G005.89$-$00.39) is a site of active star formation located 1.28~kpc away from the Sun \citep{motogi2011} that lies about 40\arcmin\ south of the W28 supernova remnant, although it is unclear if the two sources are physically related or a chance projection.  Molecular line observations give a systemic velocity of 9~km~s$^{-1}$ for W28A \citep{harvey1988,nicholas2011}.  OH$^+$ shows three distinct absorption components at about 7~km~s$^{-1}$, 13~km~s$^{-1}$, and 23~km~s$^{-1}$ (Figure \ref{fig_w28a_spectra}). The first two of these are detected in $o$-H$_2$O$^+$, but the component at  23~km~s$^{-1}$ is clearly absent.  In the two components where both ions are detected, we find molecular hydrogen fractions of about 0.085, above average in our sample.  Neither $p$-H$_2$O$^+$ nor H$_3$O$^+$ is detected in absorption toward W28A, but it is possible that the weak emission at 9~km~s$^{-1}$ in the H$_3$O$^+$ spectrum is arising in the background source itself.

\subsubsection{W31C}
Also commonly referred to as G010.62$-$00.38, W31C is an H~\textsc{ii} region within the W31 complex, and has a systemic velocity of about $-4$~km~s$^{-1}$ \citep{godard2010,gerin2010ch}.  The H~\textsc{ii} region has a large peculiar motion with respect to the Galaxy's rotation curve, and is 4.95~kpc away from the Sun as determined by H$_2$O maser observations \citep{sanna2014}.  A detailed picture of the velocity components along the line of sight is given by \citet{corbel2004}, and we use their distance estimates rather than simple kinematic rotation curve estimates in our analysis.

OH$^+$ shows absorption from about $-10$~km~s$^{-1}$ to 50~km~s$^{-1}$, and although the 971~GHz transition is saturated in multiple components the velocity profile of the 909~GHz transition is rather well matched by that of $o$-H$_2$O$^+$ (Figure \ref{fig_w31c_spectra}).  The strongest OH$^+$ and $o$-H$_2$O$^+$ absorption is in a narrow component centered at about 40~km~s$^{-1}$, which also shows absorption from H$_3$O$^+$ in both the 1655~GHz and 984~GHz lines \citep[full analysis in][]{lis2014} and $p$-H$_2$O$^+$ in the 607~GHz line. A feature in the $p$-H$_2$O$^+$ 631~GHz spectrum may also be related to this narrow component, but given the noise level we treat it as a non-detection.  A broad, weak feature in the H$_3$O$^+$ 984~GHz spectrum from about 13~km~s$^{-1}$ to 30~km~s$^{-1}$ is also thought to be caused by H$_3$O$^+$. Absorption near $-$6~km~s$^{-1}$ in the 1655~GHz H$_3$O$^+$ spectrum, however, is likely caused entirely by the 2$_{1,2}$--1$_{0,1}$ transition of H$_2^{18}$O mentioned above.

All species studied here (OH$^+$, $o$-H$_2$O$^+$, $p$-H$_2$O$^+$, and H$_3$O$^+$) were previously reported in absorption by \citet{gerin2010}.  A direct comparison of derived column densities is complicated by the different velocity intervals chosen. An analysis of the OPR of H$_2$O$^+$ toward W31C was performed by \citet{gerin2013}, and our results (see Table \ref{tbl_columns}) are in rough agreement with their findings despite the use of different velocity intervals. \citet{lis2014} also performed a multi-level analysis of H$_3$O$^+$ (using 6 transitions) in this sight line, and our reported column densities agree within uncertainties.

\subsubsection{W33A}
Trigonometric parallax observations of water masers in the W33 star forming complex put the region---including the massive young stellar object W33A---at a distance of 2.4~kpc \citep{immer2013}.  W33A (also identified as the H~\textsc{ii} region G012.90$-$00.26) has a systemic velocity of about 37~km~s$^{-1}$ as measured from various emission lines \citep[e.g.][]{vandertak2000yso,wienen2012,sanjosegarcia2013}.  OH$^+$ shows four separate absorption features from $-$4~km~s$^{-1}$ to 16~km~s$^{-1}$, 20~km~s$^{-1}$ to 25~km~s$^{-1}$, 25~km~s$^{-1}$ to 36~km~s$^{-1}$ and 36~km~s$^{-1}$ to 45~km~s$^{-1}$ (Figure \ref{fig_w33a_spectra}), all of which are also detected in $o$-H$_2$O$^+$ absorption. In all of these components we find $0.07\leq f_{\rm H_2}\leq 0.09$, above the average value for foreground gas.  Neither $p$-H$_2$O$^+$ nor H$_3$O$^+$ are detected along this sight line.

\subsubsection{G029.96$-$00.02}
The ultracompact H~\textsc{ii} region G029.96$-$00.02 is 5.26~kpc away from the Sun as determined via maser trigonometric parallax \citep{zhang2014}.  It has a systemic velocity of about 98~km~s$^{-1}$, very near the tangent velocity, and absorption occurs nearly continuously from there down to 0~km~s$^{-1}$ in several distinct velocity components, as can be seen in our OH$^+$ and $o$-H$_2$O$^+$ spectra (Figure \ref{fig_g29_spectra}).  Neither H$_3$O$^+$ nor $p$-H$_2$O$^+$ are conclusively detected, although there is a weak (2$\sigma$) feature in the 607~GHz spectrum at 71~km~s$^{-1}$ (where the strongest OH$^+$ and $o$-H$_2$O$^+$ absorption occurs) that may be due to $p$-H$_2$O$^+$.  Interestingly, three of the components along this sight line (those centered at 53~km~s$^{-1}$, 83~km~s$^{-1}$, and 92~km~s$^{-1}$) have the three lowest values of the cosmic-ray ionization rate inferred by our analysis.

\subsubsection{G034.3+00.15}
G034.3$+$00.15 shows molecular emission at about 59~km~s$^{-1}$ \citep[HCO$^+$ from][]{godard2010}, and the compact H~\textsc{ii} region is about 3.8~kpc away from the Sun as determined by a kinematic analysis \citep{fish2003}.  Absorption between about 44~km~s$^{-1}$ and 70~km~s$^{-1}$ is likely associated with the background source and molecular cloud itself, while absorption at lower velocities is due to foreground material.  The OH$^+$ and $o$-H$_2$O$^+$ spectra (Figure \ref{fig_g34_spectra}) show relatively similar absorption profiles, and H$_3$O$^+$ is not detected. The $p$-H$_2$O$^+$~607~GHz spectrum shows weak absorption at 8--16~km~s$^{-1}$ and 40--55~km~s$^{-1}$, both ranges that match the strongest OH$^+$ features.  No absorption is detected from the $p$-H$_2$O$^+$~631~GHz transition.

\subsubsection{W49N}
W49N contains several ultracompact H~\textsc{ii} regions, and at 11.11~kpc away \citep[determined from H$_2$O maser observations of][]{zhang2013} this is the most distant source we have observed.  Molecular emission peaks near 0--8~km~s$^{-1}$ for HCO$^+$ \citep{godard2010} and CH \citep{gerin2010ch}, marking the systemic velocity for W49N, and these emission features tend to be broad with FWHM$\sim$10~km~s$^{-1}$.  Absorption extends up to about 80~km~s$^{-1}$ going from the background source to the tangent point (Figure \ref{fig_w49n_spectra}), and then sweeps back down to 0~km~s$^{-1}$ going from the tangent point to the Sun \citep{fish2003}.  This means that a rotation curve analysis of the gas velocities will result in both a near and far estimate, making distance determinations highly uncertain.  Because unassociated clouds will be absorbing at the same velocities the determination of abundance ratios, $f_{\rm H_2}$, and $\zeta_{\rm H}$ will also be highly uncertain, and results from this sight line should be viewed with caution.

OH$^+$ and $o$-H$_2$O$^+$ were previously analyzed toward W49N by \citet{neufeld2010}, although only the 971~GHz OH$^+$ data were available at that time.  Column densities based on both the 909~GHz and 971~GHz transitions are similar to those reported by \citet{neufeld2010}.  The 607~GHz transition of $p$-H$_2$O$^+$ is also seen in absorption from about 35~km~s$^{-1}$ to 70~km~s$^{-1}$, and has previously been analyzed by \citet{gerin2013} for the purpose of studying the OPR of H$_2$O$^+$. Although their analysis split the H$_2$O$^+$ absorption into 5~km~s$^{-1}$ bins, the resulting OPR agree well with those we present in Table \ref{tbl_columns}.  There is a hint of absorption from the 631~GHz line of $p$-H$_2$O$^+$ near 35~km~s$^{-1}$, but interference from a strong emission line due to H$_2$CO complicates the analysis of this feature.  The 984~GHz H$_3$O$^+$ transition may show weak emission at the source velocity, but is not seen in absorption.  The 1655~GHz H$_3$O$^+$ transition potentially shows absorption (2$\sigma$ level) at 34~km~s$^{-1}$ (matches strongest OH$^+$ and H$_2$O$^+$ in velocity), but near the systemic velocity there is likely strong blending with H$_2^{18}$O absorption as was the case for W31C.

\subsubsection{W51e}
The W51 region consists of a massive molecular cloud and several active star forming complexes, and has an inferred distance of 5.41~kpc from H$_2$O maser observations \citep{sato2010}.  The compact H~\textsc{ii} regions W51~e1 and W51~e2 \citep{mehringer1994} were used as background continuum sources for our observations, and show molecular emission features centered at 55~km~s$^{-1}$ \citep{ho1996,sollins2004}.  Narrow absorption at 70~km~s$^{-1}$ is caused by a cold dense clump \citep{mookerjea2014}, while a more broadly distributed foreground cloud absorbs at 62--70~km~s$^{-1}$, and gas between about 44~km~s$^{-1}$ and 62~km~s$^{-1}$ is associated with the giant molecular cloud itself \citep{kang2010}.  Gas absorbing at lower velocities (e.g., 7~km~s$^{-1}$ and 24~km~s$^{-1}$) is well in the foreground, and likely more diffuse \citep{carpenter1998,sonnentrucker2010}.

Observations of $o$-H$_2$O$^+$ from the WISH (Water In Star-Forming regions with {\it Herschel}) program were previously presented by \citet{wyrowski2010}, and observations of $o$-H$_2$O$^+$ and OH$^+$ from the PRISMAS program by \citet{indriolo2012w51}.  Our column densities are in good agreement with those reported in the above studies, but should supersede previous values as the $o$-H$_2$O$^+$ spectra we present utilize a combination of WISH, PRISMAS, and OT1\_dneufeld\_1 data, and have significant improvement in S/N (Figure \ref{fig_w51_spectra}).  Analyses of OH$^+$ from both the 909~GHz and 971~GHz transitions are in good agreement, and differences in derived column densities can be attributed to interference from a weak emission line due to the 5$_{5,1}^+$--6$_{4,2}^+$ and 5$_{5,0}^-$--6$_{4,3}^-$ transitions of CH$_3$OH at 909.0744~GHz that can be seen in the 909~GHz spectrum as a poor fit near 85~km~s$^{-1}$.  This causes an underestimate of $N({\rm OH}^+)$ in the $42~{\rm km~s}^{-1}\leq v_{\rm LSR}\leq 55~{\rm km~s}^{-1}$ interval, and overestimate in the $-4~{\rm km~s}^{-1}\leq v_{\rm LSR}\leq 16~{\rm km~s}^{-1}$ interval.  As a result, only the 971~GHz line is used in determining $N({\rm OH}^+)$ over these intervals.

The diffuse cloud near 6~km~s$^{-1}$ shows absorption from the 607~GHz transition of $p$-H$_2$O$^+$ (absorption near 50~km~s$^{-1}$ is also likely, but interference from a strong emission line of CH$_3$OH complicates the analysis there).  Additionally, the components at 55--75~km~s$^{-1}$ are two of only four outside the Galactic center in our survey where H$_3$O$^+$ absorption is detected via the 984~GHz transition.  The features are very weak, but match exceptionally well in velocity space with absorption peaks in the $o$-H$_2$O$^+$ spectrum and both OH$^+$ spectra.  This H$_3$O$^+$ absorption denotes gas that has a high molecular fraction and is likely in a dense cloud interior rather than the diffuse outer layers \citep[following the model of][]{hollenbach2012}, a hypothesis supported by the fact that this velocity component shows the strongest absorption in HF and H$_2$O along the W51e sight line \citep{sonnentrucker2010}.  

\subsubsection{AFGL 2591}
AFGL~2591 is a cluster of high mass protostars with a bipolar outflow likely driven by the source associated with 1.3~cm and 3.6~cm continuum emission identified as VLA~3 \citep{trinidad2003}.  The molecular gas associated with the protostars has a velocity of $-5.5$~km~s$^{-1}$ \citep{vandertak1999}, and H$_2$O maser observations give a distance of 3.33~kpc \citep{rygl2012}.  OH$^+$ and H$_2$O$^+$ show two components in absorption toward AFGL~2591, at 3~km~s$^{-1}$ and $-17$~km~s$^{-1}$, neither of which matches the systemic velocity (Figure \ref{fig_afgl2591_spectra}).  The gas at 3~km~s$^{-1}$ may be associated with a foreground cloud previously reported at 0~km~s$^{-1}$ in tracers of molecular gas \citep{emprechtinger2012,vanderwiel2013}, but this requires a velocity offset between the molecular cloud and the atomic outer layers where the oxygen ions presumably reside.  The blueshifted component at $-17$~km~s$^{-1}$ may be associated with a molecular outflow \citep{mitchell1989,vandertak1999,vandertak2013H2O,vanderwiel2013}, a hypothesis that could explain the larger value of $f_{\rm H_2}=0.09$ found in this component. Both OH$^+$ and H$_2$O$^+$ have previously been studied toward AFGL~2591 \citep{bruderer2010,benz2013} along with several other light hydrides.  The column densities that we derive for the two velocity components are in relatively good agreement with those found by \citet{bruderer2010}, as well as the line of sight column densities reported by \citet{benz2013}.  H$_3$O$^+$ absorption is not detected, likely due to the low continuum level signal-to-noise ratio, although emission from the $4_{3}^{+}-3_{3}^{-}$ transition has been observed at the systemic velocity \citep{benz2013}. A more detailed study of light hydrides in AFGL~2591 is currently underway (Benz et al. 2015, in preparation).

\subsubsection{DR21C and DR21(OH)}
DR21C and DR21(OH) are compact H~\textsc{ii} regions that are parts of the DR21 molecular ridge, a region of massive star formation about 1.5~kpc away from the Sun \citep[determined from H$_2$O maser observations by][]{rygl2012}. Systemic velocities for both sources are about $-3$~km~s$^{-1}$ \citep{vandertak2010,zapata2012}. The sources are separated by 3.1\arcmin\ on sky, corresponding to a projected separation of 1.3~pc at the adopted distance. The OH$^+$ and $o$-H$_2$O$^+$ absorption profiles for DR21C and DR21(OH) are largely similar (Figures \ref{fig_dr21_spectra} and \ref{fig_dr21oh_spectra}).  In the OH$^+$ spectra there is a shallow absorption wing from about 25~km~s$^{-1}$ to 15~km~s$^{-1}$, followed by a rapid increase to maximum absorption near 9~km~s$^{-1}$.  The absorption then gradually decreases until it disappears around $-15$~km~s$^{-1}$.  The most notable difference between the spectra is that toward DR21C there is a local minimum in absorption at $-5$~km~s$^{-1}$, while for DR21(OH) the absorption decreases monotonically below 0~km~s$^{-1}$.  Spectra of $o$-H$_2$O$^+$ show the same general structure.  Differences in the absorption profiles between the two sources only occur near the systemic velocities, suggesting that most of the absorption arises in a common foreground cloud.  Indeed, the strongest absorption at 9~km~s$^{-1}$ matches a foreground cloud observed in CO and HCO$^+$ associated with the nearby source W75N \citep{schneider2010}.  Neither the 607~GHz nor the 631~GHz transition of $p$-H$_2$O$^+$ is detected toward DR21(OH), but the 607~GHz line shows absorption toward DR21C, although there is likely interference from emission lines of CH$_3$OH and H$^{13}$CO$^+$. H$_3$O$^+$ is not detected in either sight line.  Previous observations of $o$-H$_2$O$^+$ toward DR21C were reported by \citet{ossenkopf2010}, and our resulting column densities are in relatively good agreement.  Our inferred column densities for OH$^+$ and H$_2$O$^+$ are also in good agreement with those found as part of a more detailed analysis of the DR21C sight line (Chambers et al. 2015, in preparation)

\subsubsection{NGC~7538 IRS1}
The hyper-compact H~\textsc{ii} region NGC~7538~IRS1 is 2.65~kpc distant as determined via trigonometric parallax of CH$_3$OH masers \citep{moscadelli2009} and has a systemic velocity of $-59$~km~s$^{-1}$ observed in several molecules \citep{zhu2013}.  OH$^+$ and $o$-H$_2$O$^+$ show very similar absorption profiles with components at $-50$~km~s$^{-1}$, $-33$~km~s$^{-1}$, $-28$~km~s$^{-1}$, $-7$~km~s$^{-1}$, and 0~km~s$^{-1}$, the exception being that the $-$28~km~s$^{-1}$ component is missing in $o$-H$_2$O$^+$ (Figure \ref{fig_ngc7538_spectra}).  Absorption from $-$60~km~s$^{-1}$ to $-$40~km~s$^{-1}$ is likely associated with material at the background source, while the other components arise in foreground gas.  A detailed analysis of light hydrides in NGC~7538~IRS1 is forthcoming in Benz et al. 2015 (in preparation).

\subsubsection{W3~IRS5 and W3(OH)}
Both W3(OH) and W3~IRS5 are located in the W3 molecular cloud complex, a site of active star formation within the Galaxy.  W3(OH) is an ultracompact H~\textsc{ii} region thought to harbor a massive young star, while W3~IRS5 is a protocluster of a few high mass stars.  Multi-epoch VLBA observations of water masers toward both sources have been used to determine distances of $2.04\pm0.07$~kpc \citep{hachisuka2006} and $1.83\pm0.14$~kpc \citep{imai2000} for  W3(OH) and W3~IRS5, respectively.  Molecular line observations show systemic velocities of $-46$~km~s$^{-1}$ for W3(OH) \citep{wilson1991} and $-39$~km~s$^{-1}$ for W3~IRS5 \citep{wang2013}. The two sources are 16.6\arcmin\ apart in the sky, corresponding to a projected separation of about 9.7~pc at the distance of the background sources.

The absorption profiles of OH$^+$ and $o$-H$_2$O$^+$ toward W3(OH) and W3~IRS5 are largely similar, with absorption from about 7~km~s$^{-1}$ to $-27$~km~s$^{-1}$ (Figures \ref{fig_w3irs5_spectra} and \ref{fig_w3oh_spectra}). These features are due to foreground clouds that have previously been observed in H~\textsc{i} absorption at about 0~km~s$^{-1}$ and $-20$~km~s$^{-1}$, and which are estimated to be at distances of 0.7~kpc and 1.5~kpc, respectively \citep{normandeau1999}.  Toward W3(OH) the absorption between $-41$~km~s$^{-1}$ and $-50$~km~s$^{-1}$ is likely associated with material surrounding the background source itself, and similar for the $-37$~km~s$^{-1}$ to $-47$~km~s$^{-1}$ absorption toward W3~IRS5.  Neither line of sight shows a conclusive detection of the $p$-H$_2$O$^+$ line at 607~GHz, nor is H$_3$O$^+$ detected toward W3(OH).  Emission features in the 909~GHz spectrum toward W3(OH) and in the 971~GHz spectrum toward W3~IRS5 are due to CH$_3$OH.

Analyses of light hydrides in the W3~IRS5 sight line have been previously reported by \citet{benz2010,benz2013}.  Column densities that we find for OH$^+$ and $o$-H$_2$O$^+$ in the foreground gas are consistent with those reported by \citet{benz2010}, within uncertainties.  For the gas associated with the background source, however, our column densities are about half of the values reported in Benz et al. 2015 (in preparation).  The difference arises because we assume the entire populations of both species are in the ground state, while Benz et al. 2015 (in preparation) adopt a higher excitation temperature to account for heating by UV radiation, assuming the gas is located in the cavity wall of a protostellar outflow.

\subsubsection{G327.3-0.6}
G327.3$-$0.6 is a hot core within a region of active star formation, and provides the longest line of sight probing the fourth Galactic quadrant in our study.  Molecular emission lines give a systemic velocity of $-$44.5~km~s$^{-1}$ \citep{sanjosegarcia2013,leurini2013}, and a distance of 3.3~kpc was determined via a kinematic analysis of H~\textsc{i} absorption data \citep{urquhart2012}.  OH$^+$ and $o$-H$_2$O$^+$ show similar absorption profiles for the most part (Figure \ref{fig_g327_spectra}). Absorption from $-$31~km~s$^{-1}$ to $-$55~km~s$^{-1}$ likely arises within the cloud containing the background source, while features at $v_{\rm LSR}\geq-26$~km~s$^{-1}$ are caused by foreground material.  It is unclear why the expected absorption due to the weakest hyperfine component of the OH$^+$ 971~GHz transition fails to match the observed spectrum near $-$75~km~s$^{-1}$.

\subsubsection{NGC 6334 I and NGC 6334 I(N)}
The sources NGC~6334~I (a hot molecular core) and NGC~6334~I(N) (a mid-IR quiet high mass protostellar object) are both within the NGC~6334 complex of molecular clouds and H~\textsc{ii} regions, located at a distance of 1.35~kpc \citep{wu2014}. Systemic velocities for the two sources are $-$7.7~km~s$^{-1}$ and $-$4.5~km~s$^{-1}$, respectively, and they are separated by 1.9\arcmin\ on sky, corresponding to projected separation of 0.74~pc at the adopted distance.  Absorption profiles of OH$^+$ and $o$-H$_2$O$^+$ are nearly identical between the two sight lines with peaks at 3~km~s$^{-1}$ and $-$2~km~s$^{-1}$, although NGC~6334~I also shows a weaker component near $-$10~km~s$^{-1}$ (Figures \ref{fig_ngc6334I_spectra} and \ref{fig_ngc6334In_spectra}).  None of these components are well-matched to those seen in H$_2$O \citep{vandertak2013H2O} and HF \citep{emprechtinger2012} that have been attributed to protostellar envelopes, outflows, and foreground clouds, further highlighting the different regions traced by such molecules.  Given the similarities between the two sight lines though, we can conclude that the absorption features at 3~km~s$^{-1}$ and $-$2~km~s$^{-1}$ likely arise in a common foreground cloud. A detailed analysis of light hydrides in the background sources will be presented by Benz et al. 2015 (in preparation).  Column densities of OH$^+$ and H$_2$O$^+$ toward NGC~6334~I have previously been reported by \citet{zernickel2012}.  Our findings for OH$^+$ where they adopt an excitation temperature of 2.7~K are in good agreement, but for H$_2$O$^+$ their adopted value of $T_{ex}=24$~K leads to a much larger column density.

\subsection{H$_2$O$^+$ ortho-to-para ratio (OPR)}
Out of our entire survey, 11 velocity intervals show conclusive detections of the $p$-H$_2$O$^+$ line at 607~GHz, and only 6 of those are above a 3$\sigma$ level.  Four of the detections are along Galactic center sight lines, including one toward M$-$0.02$-$0.07 and three toward Sgr~B2(M) previously reported by \citet{schilke2013}.  These 11 components provide the opportunity to investigate the OPR of H$_2$O$^+$, which is given in Table \ref{tbl_columns} column 9.  In all cases, within uncertainties the OPR is consistent with a value of 3, the ratio expected in the high temperature limit based solely on nuclear spin statistical weights.  While it is possible that reactive collisions, temperature, forbidden spontaneous emission \citep{tanaka2013}, and state-specific formation and destruction can skew the OPR away from 3, observations thus far have not conclusively demonstrated any such deviations in the diffuse ISM \citep{gerin2013,schilke2013}.

\subsection{H$_3$O$^+$ Detections}
H$_3$O$^+$ is only detected in absorption in 16 components, 12 of which are in sight lines toward the Galactic center.  Models of the chemistry surrounding oxygen-bearing ions find that H$_3$O$^+$ will only form in observable abundances in gas that is well shielded from the interstellar radiation field \citep[visual extinction, $A_V\gtrsim 3$~mag;][]{hollenbach2012}.  In such regions, the oxygen chemistry is driven by the reaction $\mathrm{O+H_3^+\rightarrow OH^{+}+H_{2}}$ rather than reaction (\ref{reac_Op_H2}), so abundances are linked to the ionization rate of H$_2$ instead of H.  The small number of H$_3$O$^+$ detections in our sample suggests that most of the components we consider have low $A_V$, and are comprised of diffuse gas.  This supports the use of diffuse cloud chemistry in our analysis, the link between OH$^+$ and the ionization rate of atomic hydrogen, and the assumption that molecules are almost entirely in their respective ground states.

\subsection{Molecular hydrogen fraction and OH$^+$ to H$_2$O$^+$ ratio}

\begin{figure}
\epsscale{0.8}
\plotone{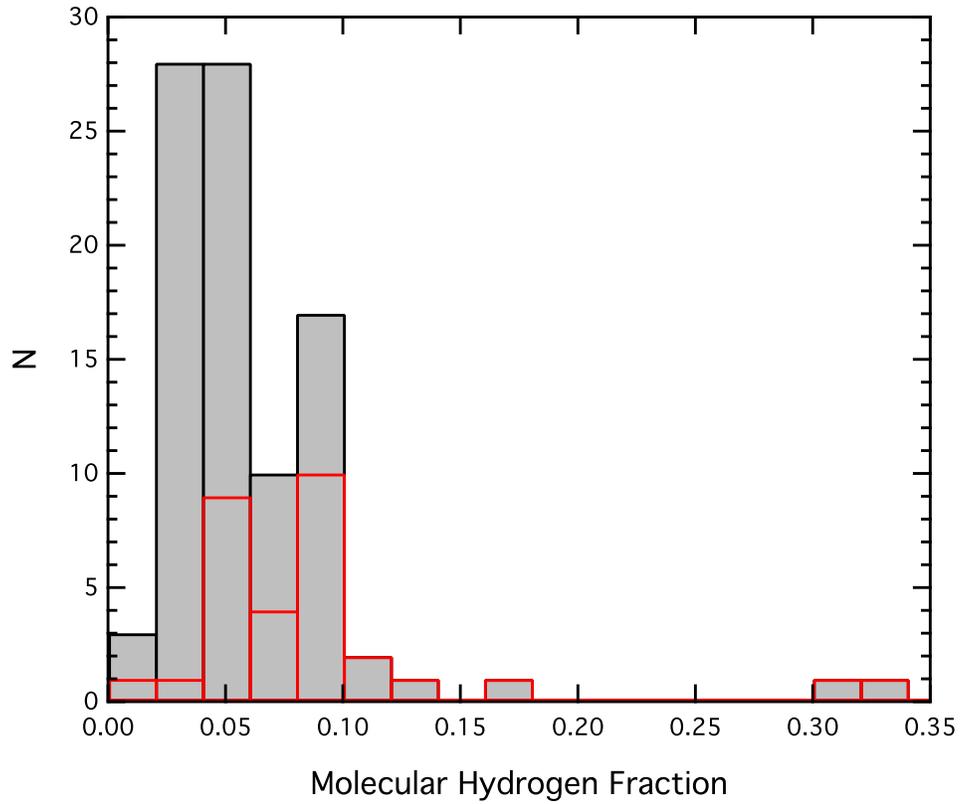}
\caption{Distribution of $f_{\rm H_2}$ as determined from our analysis of OH$^+$ and H$_2$O$^+$ abundances. The filled gray bars show all velocity intervals where $f_{\rm H_2}$ is computed, and the red bars mark the distribution for velocity intervals within 5~km~s$^{-1}$ of the systemic velocity of the background source (i.e., that may be associated with material surrounding the continuum source).}
\label{fig_fh2_hist}
\end{figure}

The abundance ratio $N({\rm OH}^+)/N({\rm H_{2}O}^+)$ is inversely related to $f_{\rm H_2}$, as clearly seen in equation (\ref{eq_hydride_fH2}). Conceptually this is easy to understand as more H$_2$ will drive the OH$^+$ + H$_2$ reaction more rapidly, converting more OH$^+$ into H$_2$O$^+$.  Values of $N({\rm OH}^+)/N({\rm H_{2}O}^+)$ and $f_{\rm H_2}$ are given in columns 5 and 9 of Table \ref{tbl_results}, respectively, and the distribution of $f_{\rm H_2}$ is presented as a histogram in Figure \ref{fig_fh2_hist}.  The observed OH$^+$ and H$_2$O$^+$ abundances favor gas with low molecular hydrogen fractions, as all but three of the components in the Galactic disk have $f_{\rm H_2}<0.1$, and only in the Galactic center does $f_{\rm H_2}$ exceed 0.15. Excluding data in the Galactic center sight lines, the distribution of molecular hydrogen fractions in our sample has mean 0.053 and standard deviation 0.026.  

We have also considered whether or not $f_{\rm H_2}$ differs in velocity intervals that are potentially associated with material surrounding our target background sources (contain absorption within 5~km~s$^{-1}$ of systemic velocity).  The distribution of $f_{\rm H_2}$ in these components is shown by the red bars in Figure \ref{fig_fh2_hist}, and it is clear that they tend to have larger molecular hydrogen fractions. If these components potentially associated with background sources are also excluded from our analysis, the mean and standard deviation of $f_{\rm H_2}$ in our sample change to 0.042$\pm$0.018. These results are consistent with the findings of previous studies utilizing OH$^+$ and H$_2$O$^+$ observations for the same purpose \citep{gerin2010,neufeld2010,indriolo2012w51,vandertak2013}, as well as with models of oxygen chemistry \citep{hollenbach2012}, confirming the trend that the two species predominantly reside in mostly atomic gas.

\begin{figure}
\epsscale{0.7}
\plotone{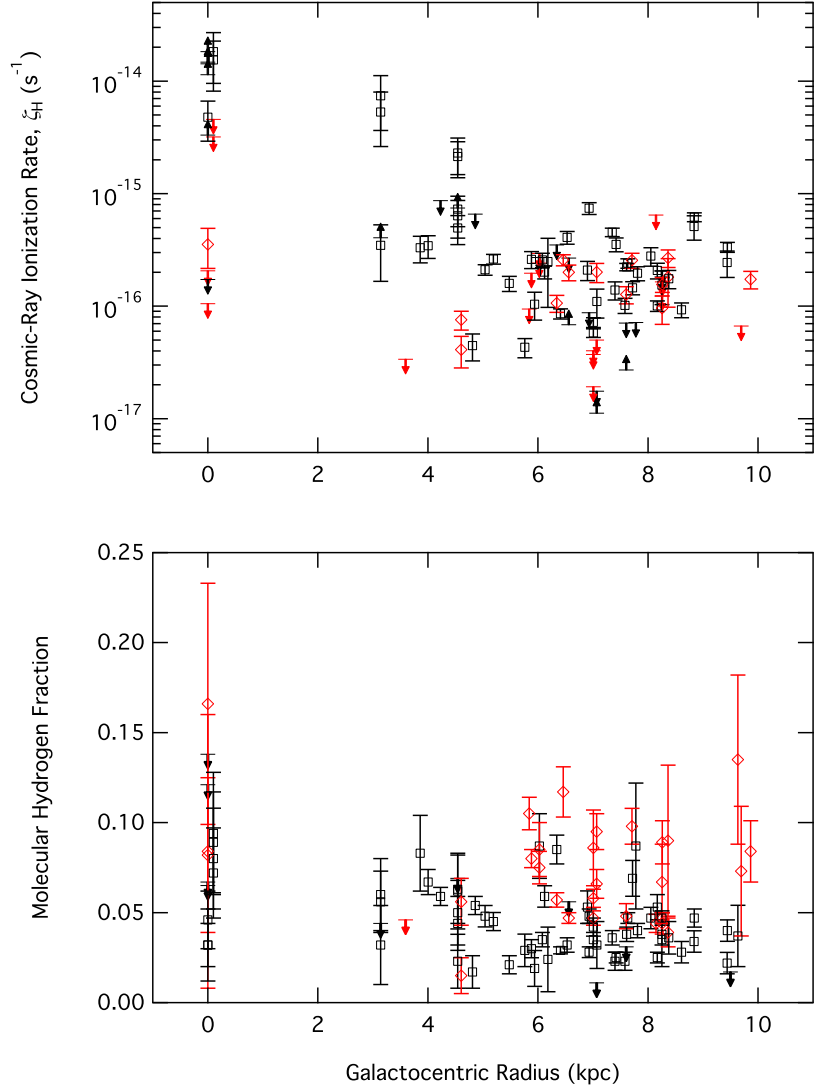}
\caption{Top: Cosmic-ray ionization rate versus Galactocentric radius. Bottom: Molecular hydrogen fraction versus Galactocentric radius.  Red diamonds denote velocity intervals within 5~km~s$^{-1}$ of the systemic velocity of the background source.  Black squares denote foreground clouds. Upper limits and lower limits are marked by arrows, and use the same color scheme denoting foreground versus background. Note there are 4 components, all in the Galactic center, with $f_{\rm H_2}>0.25$, but we have scaled the axis to more clearly show the entire data set.}
\label{fig_GCrad_fH2_zetaH}
\end{figure}

A plot of the molecular hydrogen fraction versus Galactocentric radius is shown in the bottom panel of Figure \ref{fig_GCrad_fH2_zetaH}. Red diamonds denote velocity intervals more likely associated with background sources and black squares those thought to be foreground clouds, and there is distinct separation between the bulk of the two samples as would be expected given the discussion above.  There does not appear to be any relation between $f_{\rm H_2}$ and $R_{\rm gal}$, either for the entire sample or for the sub-samples separately.  If metallicity increases toward the Galactic center though \citep[][and references therein]{wolfire2003}, $x_e$ should as well, and larger values of $f_{\rm H_2}$ would be required to produce the observed $N({\rm OH}^+)/N({\rm H_{2}O}^+)$ ratios.  Whether or not $f_{\rm H_2}$ changes with $R_{\rm gal}$ then hinges on the underlying assumption that $x_e$ is either constant or variable with Galactocentric radius.

\subsection{Cosmic-ray ionization rate}

\begin{figure}
\epsscale{0.8}
\plotone{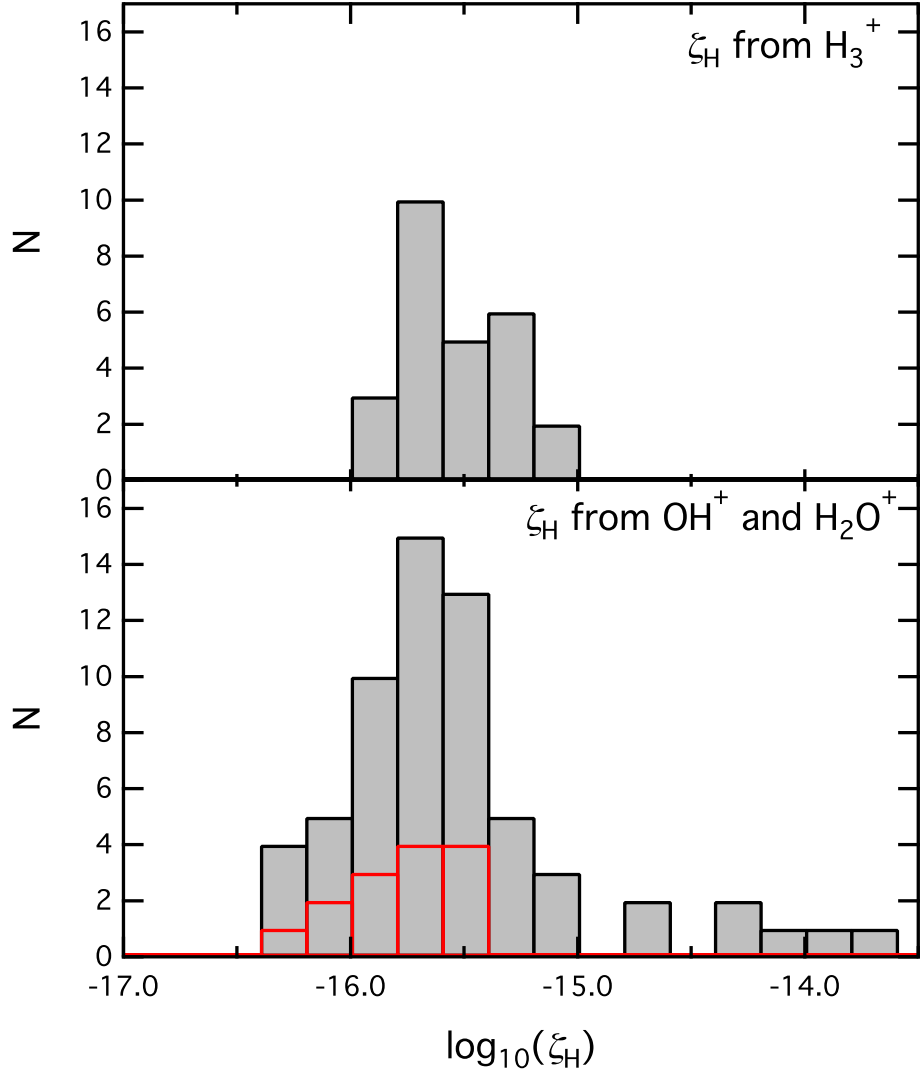}
\caption{Histogram of $\zeta_{\rm H}$ as determined from abundances of OH$^+$ and H$_2$O$^+$ (bottom panel) and from H$_3^+$ in diffuse clouds \citep[top panel;][and unpublished data]{indriolo2012}. In the bottom panel filled gray bars show all velocity intervals where $\zeta_{\rm H}$ is computed, and the red bars mark the distribution for velocity intervals within 5~km~s$^{-1}$ of the systemic velocity of the background source (i.e., that may be associated with material surrounding the continuum source). In the top panel, only diffuse cloud sight lines where H$_3^+$ is detected have been used in creating the histogram of ionization rates. Over half of all sight lines observed searching for H$_3^+$ resulted in non-detections; upper limits on the ionization rate range from a few times 10$^{-17}$~s$^{-1}$ up to 10$^{-15}$~s$^{-1}$.}
\label{fig_zetaH_hist}
\end{figure}

The final column of Table \ref{tbl_results} gives the cosmic-ray ionization rates inferred from our analysis, and the distribution of $\zeta_{\rm H}$ is presented in the bottom panel of Figure \ref{fig_zetaH_hist}.  Upper limits on $\zeta_{\rm H}$ are the result of optically thick 21~cm H~\textsc{i} absorption that only allows us to place lower limits on $N({\rm H})$.  Lower limits on $\zeta_{\rm H}$ arise when we are only able to place a lower limit on $N({\rm OH}^+)$.  A range of ionization rates is reported when H$_2$O$^+$ is not detected, with the upper bound determined by the upper limit on $N({\rm H_2O^+})$, and the lower bound determined in the limit where $N({\rm H_2O^+}) \rightarrow 0$.  Uncertainties in $\zeta_{\rm H}$ only account for the uncertainties in  observed column densities, and do not include the effects discussed in Section \ref{subsec_unc}.  As before, in Figure \ref{fig_zetaH_hist} the grey bars represent the total sample of velocity intervals where the ionization rate has been determined, and the red bars denote the sub-sample of clouds that may be associated with background sources.  All components with $\zeta_{\rm H}>10^{-15}$~s$^{-1}$ arise in sight lines toward the Galactic center, and due to the unique nature of this region we exclude all data from the M$-$0.13$-$0.08, M$-$0.02$-$0.07, Sgr~B2(M), and Sgr~B2(N) sight lines during the following analysis. 

The distribution of ionization rates inferred from OH$^+$ and H$_2$O$^+$ appears to be log-normal. We find the mean value of $\log(\zeta_{\rm H})$ to be -15.75 ($\zeta_{\rm H}=1.78\times10^{-16}$~s$^{-1}$) with standard deviation 0.29. The distribution in components potentially associated with background sources does not differ appreciably, although it lacks some of the highest ionization rates seen in the foreground clouds. Shown in the top panel of Figure \ref{fig_zetaH_hist} is the distribution of ionization rates in diffuse molecular clouds found by \citet{indriolo2012} using observations of H$_3^+$. Ionization rates of molecular hydrogen ($\zeta_2$) reported therein have been scaled by 1.5/2.3 to convert to the ionization rate of atomic hydrogen \citep{glassgold1973,glassgold1974}. This sample has a mean value of -15.55 ($\zeta_{\rm H}=2.82\times10^{-16}$~s$^{-1}$) and standard deviation 0.24.  Despite slight differences, mean ionization rates calculated using the different molecules are in agreement.  To check whether or not the two distributions of ionization rates differ, we performed a two-sample K-S test, and we cannot reject the hypothesis that the two samples are drawn from the same underlying distribution.  The greatest difference in the two distributions occurs for $\zeta_{\rm H}\lesssim1.5\times10^{-16}$~s$^{-1}$, and no ionization rates inferred from H$_3^+$ are below $10^{-16}$~s$^{-1}$.  Likely this is because H$_3^+$ absorption lines are fairly weak (only a few percent deep at most), and at low ionization rates the molecule will not be produced in detectable abundances.  This means OH$^+$ and H$_2$O$^+$ are important tracers of $\zeta_{\rm H}$ in a regime where H$_3^+$ is unobservable.

Cosmic-ray ionization rate versus Galactocentric radius is shown in the top panel of Figure \ref{fig_GCrad_fH2_zetaH}.  Outside a radius of 5~kpc there does not seem to be any relation between $\zeta_{\rm H}$ and $R_{\rm gal}$.  This appears to agree with the conclusion of a uniform cosmic-ray density drawn from gamma-ray observations tracing the flux of $E\gtrsim300$~MeV protons \citep{ackermann2011}.  Within the Galactic center itself there is a large range of ionization rates, including six components with $\zeta_{\rm H}>10^{-14}$~s$^{-1}$. These are the highest values found in our study, and they all come from gas toward M$-$0.13$-$0.08 and M$-$0.02$-$0.07 with $-$159~km~s$^{-1}\leq v_{\rm LSR}\leq -$85~km~s$^{-1}$ and toward Sgr~B2(M) and Sgr~B2(N) with $-$130~km~s$^{-1}\leq v_{\rm LSR}\leq -$60~km~s$^{-1}$. OH$^+$ shows continuous, substantial absorption over these velocities (see Figures \ref{fig_sgra20_spectra}--\ref{fig_sgrb2n_spectra}), while H~\textsc{i} only has minimal absorption in the same range \citep[Figure 7 in][Figure 5 position 7]{lang2010,dwarakanath2004}. As mentioned above, at such high ionization rates equation (\ref{eq_zeta_n}) is no longer a valid approximation because electrons freed during the ionization of H and H$_2$ make $x_e$ strongly dependent on $\zeta_{\rm H}$.  Because our adopted value of $x_e$ is likely an underestimate, the high ionization rates reported in the Galactic center should still be valid lower limits. Smaller ionization rates in the Galactic center are found in the velocity intervals corresponding to all four of the background sources---regions known to be largely molecular.  Indeed, the strong H$_3$O$^+$ absorption in these components requires large H$_2$ abundances and denser gas.  The diffuse cloud chemistry used to infer $f_{\rm H_2}$ and $\zeta_{\rm H}$ is almost assuredly not valid in these regions, and the higher ionization rates found in other components will be more indicative of the particle flux in the Galactic center. Previous studies of the Galactic center region also find cosmic-ray ionization rates on the order of $10^{-15}$--$10^{-13}$~s$^{-1}$. Observations of H$_3^+$ show the molecule to be widespread in the CMZ, and inferred ionization rates are several times $10^{-15}$~s$^{-1}$ on average \citep{oka2005,goto2008}.  Analysis of the 6.4~keV Fe K$\alpha$ line, gamma rays, and radio synchrotron emission in the Galactic center also points to a large population of energetic particles, and estimates of the resulting ionization rate range from a few times 10$^{-15}$~s$^{-1}$ up to $5\times10^{-13}$~s$^{-1}$ depending on the location in question \citep{yusefzadeh2007,yusefzadeh2013}.  

Sight lines toward the Galactic center also show OH$^+$ and H$_2$O$^+$ absorption from the 3~kpc and 4.5~kpc spiral arms.  Ionization rates in these components tend to be higher than most of those found at larger $R_{\rm gal}$, and lower than those found in the Galactic center, indicative of a gradient in $\zeta_{\rm H}$.  Such a gradient was predicted by \citet{wolfire2003}, and is expected given the high concentration of energetic sources in the inner Galaxy leads to more particle acceleration than elsewhere in the disk. However, we must re-emphasize that absorption attributed to these spiral arms is very likely blended with absorption from gas within the CMZ, so it is possible that the intermediate ionization rates are simply a combination of high ionization rates in the Galactic center and average ionization rates in the spiral arms. Additional observations at $R_{\rm gal}\leq5$~kpc are necessary to distinguish between the two interpretations.

Another relationship that has been the focus of recent studies is that between $\zeta_{\rm H}$ and $N_{\rm H}\equiv N({\rm H})+2N({\rm H}_2)$, the total column density of a given cloud.  The cross section for ionization of H and H$_2$ by cosmic rays increases with decreasing energy, meaning the flux of low-energy particles ($E\leq100$~MeV) is most important in controlling $\zeta_{\rm H}$, and such particles will quickly be removed from the cosmic-ray spectrum due to these energy losses \citep[e.g.,][]{padovani2009}.  Cosmic-ray ranges (expressed as the product of density and distance, i.e., column density, through which a particle can propagate before losing all of its energy to ionization interactions) have been calculated as a function of particle energy and are available via a NIST web query.\footnote{http://www.nist.gov/pml/data/star/index.cfm} Ranges for 1~MeV, 10~MeV, and 100~MeV protons propagating through a gas of purely atomic hydrogen are $Rn({\rm H})=5.1\times10^{20}$~cm$^{-2}$, $3.2\times10^{22}$~cm$^{-2}$, and $2.2\times10^{24}$~cm$^{-2}$, respectively. Given these ranges and an average diffuse cloud with $N_{\rm H}=10^{21}$~cm$^{-2}$, the higher-energy particles will pass through the entirety of the cloud, while the lower energy particles---those most important for ionization---will be stopped part of the way through the cloud.  The expected result then, is that $\zeta_{\rm H}$ will decrease with increasing $N_{\rm H}$ as the particles most efficient at ionization are removed from the spectrum.

\begin{figure}
\epsscale{1.0}
\plotone{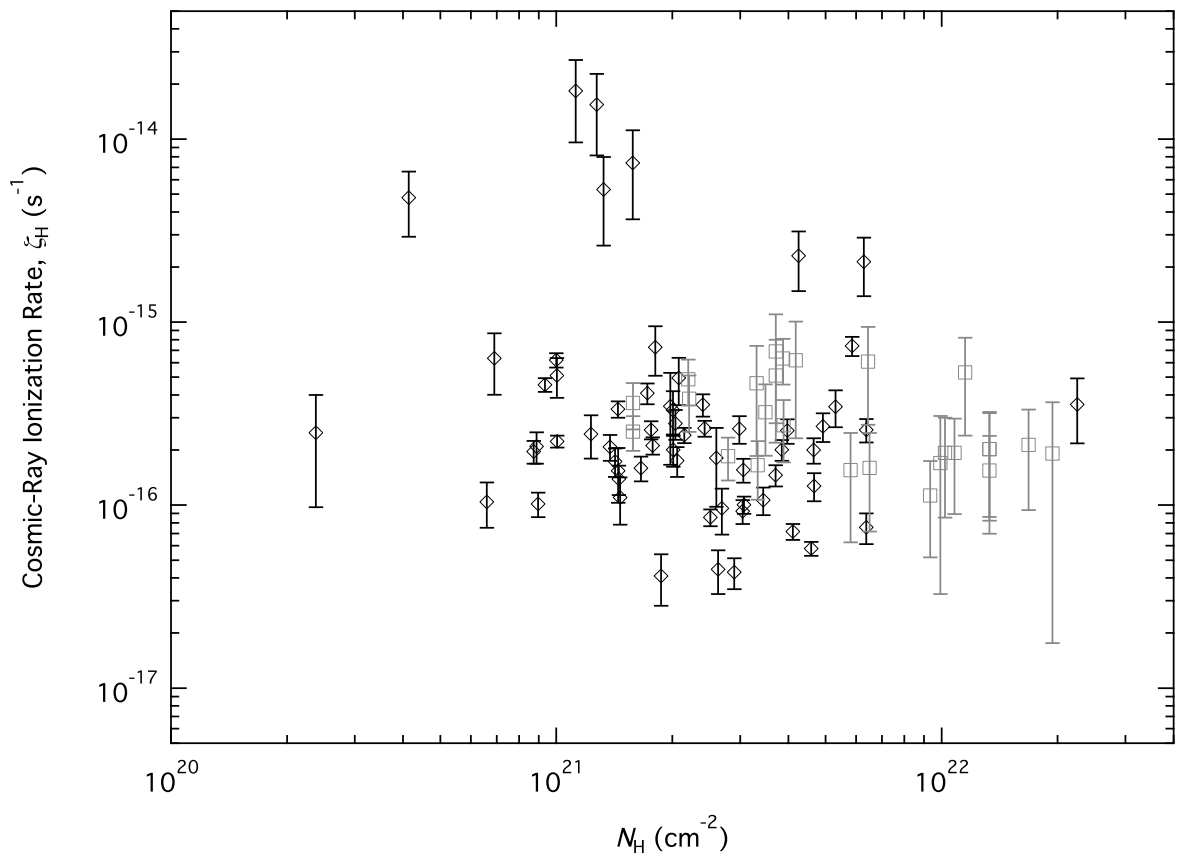}
\caption{Cosmic-ray ionization rate versus total hydrogen column density, $N_{\rm H}$.  We have estimated $N_{\rm H}$ using values of $N({\rm H})$ and $f_{\rm H_2}$ reported in Table \ref{tbl_results}.  Black diamonds are from the present study, and grey squares from H$_3^+$ observation of \citet{indriolo2012}. All ionization rates above $10^{-15}$~s$^{-1}$ are from sight lines toward the Galactic center.}
\label{fig_zetaH_NH}
\end{figure}

In Figure \ref{fig_zetaH_NH} we plot $\zeta_{\rm H}$ versus $N_{\rm H}$ for the sample studied herein, and for ionization rates determined from H$_3^+$ observations \citep{indriolo2012}.  We see no change in $\zeta_{\rm H}$ over the range $N_{\rm H}=0.7$--20$\times10^{21}$~cm$^{-2}$, consistent with our previous findings.  Only for clouds with $N_{\rm H}\gtrsim10^{23}$~cm$^{-2}$ do reported ionization rates decrease significantly \citep[e.g., see][and references therein]{padovani2009}, hinting at the loss of low-energy cosmic rays.  The lack of a correlation between $\zeta_{\rm H}$ and $N_{\rm H}$ in diffuse clouds may be due to multiple effects.  Even if the column density along a line of sight is large enough to stop low-energy particles, it is possible that the amount of material a particle would have traverse to reach that point moving in the plane of the sky is much lower.  It is also possible that what appears as a single absorption feature in velocity space is actually composed of several discrete clouds along the line of sight, each with column densities much smaller than the total.  Finally, due to the small molecular hydrogen fractions we have concluded that OH$^+$ and H$_2$O$^+$ reside predominantly in the outer layers of clouds.  This means that our inferred ionization rates are based on material expected to experience a mostly unattenuated flux of low-energy cosmic rays.

\section{SUMMARY} \label{sec_summary}
We have surveyed 20 sight lines in the Galactic disk with the {\it Herschel Space Observatory}, all of which show absorption from OH$^+$ and $o$-H$_2$O$^+$.  Sight lines have been sub-divided by velocity intervals into a total sample of 105 components where we determine column densities for the observed species.  H$_3$O$^+$ is detected in only 4 components outside of the Galactic center, suggesting the majority of the gas being probed is diffuse and at $A_V\lesssim3$~mag.  Abundances are used to infer both the molecular hydrogen fraction and cosmic-ray ionization rate in each component.  The vast majority of components have $f_{\rm H_2}\leq0.1$, confirming previous findings that OH$^+$ and H$_2$O$^+$ reside in primarily atomic gas, likely in the outer layers of clouds.  We find a distinct difference in the distribution of $f_{\rm H_2}$ in foreground components versus the distribution in components potentially associated with material surrounding background sources (i.e., envelopes, outflows), with the latter showing larger molecular hydrogen fractions.  The distribution of $f_{\rm H_2}$ in foreground components is described by a Gaussian function with mean and standard deviation $0.042\pm0.018$. We find no correlation between molecular hydrogen fraction and Galactocentric radius, although this is dependent on the assumption of a constant $x_e$.  If the electron fraction varies with $R_{\rm gal}$ (perhaps in unison with the known metallicity gradient), then $f_{\rm H_2}$ would increase toward the Galactic center.

Our study has more than doubled the sample of Galactic diffuse molecular clouds where the cosmic-ray ionization rate has been determined. Ionization rates inferred from OH$^+$ and H$_2$O$^+$ outside the Galactic center show a log-normal distribution with mean -15.75 ($\zeta_{\rm H}=1.78\times10^{-16}$~s$^{-1}$) and standard deviation 0.29.  This distribution is consistent with that found using H$_3^+$ observations along diffuse molecular cloud sight lines, and the mean ionization rates found using the different molecular tracers agree within uncertainties. Given these results and the size of our sample, we confirm the findings that average cosmic-ray ionization rates in the Galactic disk are on the order of $10^{-16}$~s$^{-1}$.

Cosmic-ray ionization rates in the Galactic center are 1--2 orders of magnitude larger than those found in the Galactic disk, again consistent with previous findings.  It is possible that there is a gradient in $\zeta_{\rm H}$, with the ionization rate decreasing from the Galactic center out to $R_{\rm gal}\approx5$~kpc, but for $R_{\rm gal}>5$~kpc $\zeta_{\rm H}$ shows no correlation with Galactocentric radius.  This is in agreement with the gamma-ray signature from $E\geq300$~MeV protons interacting with ambient gas, and it is interesting that particles at these different energies show similar behavior despite significantly different ranges.

\mbox{}
Support for this work was provided by NASA through an award issued by JPL/Caltech. N.I. and D.A.N. are funded by NASA Research Support Agreement No. 1393741 provided through JPL. J.R.G. thanks the Spanish MINECO for funding support under grants CSD2009-00038 and AYA2012- 32032. The authors thank Vincent Fish for providing digital copies of H~\textsc{i} spectra from his 2003 paper, and the anonymous referee for insightful comments and suggestions. HIFI has been designed and built by a consortium of institutes and university departments from across Europe, Canada and the United States under the leadership of SRON Netherlands Institute for Space Research, Groningen, The Netherlands and with major contributions from Germany, France and the US. Consortium members are: Canada: CSA, U.Waterloo; France: CESR, LAB, LERMA, IRAM; Germany: KOSMA, MPIfR, MPS; Ireland, NUI Maynooth; Italy: ASI, IFSI-INAF, Osservatorio Astrofisico di Arcetri-INAF; Netherlands: SRON, TUD; Poland: CAMK, CBK; Spain: Observatorio Astronómico Nacional (IGN), Centro de Astrobiología (CSIC-INTA). Sweden: Chalmers University of Technology - MC2, RSS \& GARD; Onsala Space Observatory; Swedish National Space Board, Stockholm University - Stockholm Observatory; Switzerland: ETH Zurich, FHNW; USA: Caltech, JPL, NHSC.



\clearpage

\clearpage

\clearpage

\clearpage

\clearpage

\clearpage

\clearpage

\clearpage

\clearpage

\clearpage

\clearpage

\clearpage

\clearpage

\clearpage

\clearpage

\clearpage

\clearpage

\clearpage

\clearpage

\clearpage

\clearpage

\clearpage

\clearpage

\clearpage

\clearpage


\begin{deluxetable}{lccccccccc}
\tablewidth{0pt}
\tabletypesize{\scriptsize}
\tablecaption{Targeted Transitions\label{tbl_transitions}} 
\tablehead{Molecule &  \multicolumn{3}{c}{Transition}   & Rest Frequency & $E_{l}/k$ & $g_u$ & $g_l$ & $A$  \\
& & & & (MHz) & (K) & & & (10$^{-2}$~s$^{-1}$)}
\startdata
  & $N'$--$N''$ & $J'$--$J''$ & $F'$--$F''$ &  &  &  &  &  \\ 
OH$^+$ & 1--0 & 0--1 & 1/2--1/2 & \phantom{1}909045.2 & 0.0055 & 2 & 2 & 0.52 \\
OH$^+$ & 1--0 & 0--1 & 1/2--3/2 & \phantom{1}909158.8\tablenotemark{a} & 0 & 2 & 4 & 1.05 \\
\hline
OH$^+$ & 1--0 & 2--1 & 5/2--3/2 & \phantom{1}971803.8\tablenotemark{a} & 0 & 6 & 4 & 1.82 \\
OH$^+$ & 1--0 & 2--1 & 3/2--1/2 & \phantom{1}971805.3 & 0.0055 & 4 & 2 & 1.52 \\
OH$^+$ & 1--0 & 2--1 & 3/2--3/2 & \phantom{1}971919.2 & 0 & 4 & 4 & 0.30 \\
\hline
OH$^+$ & 1--0 & 1--1 & 1/2--1/2 & 1032997.9 & 0.0055 & 2 & 2 & 1.41 \\
OH$^+$ & 1--0 & 1--1 & 3/2--1/2 & 1033004.4 & 0.0055 & 4 & 2 & 0.35 \\
\enddata
\tablecomments{All data were obtained from the Cologne Database for Molecular Spectroscopy \citep[CDMS;][]{muller2005}, although some frequencies have been updated as described below. OH$^+$ data are from \citet{bekooy1985}.  H$_2$O$^+$ data are from \citet{murtz1998}, but the frequencies for the transitions at 1115~GHz have been shifted by +5~MHz.  This shift provides the best match between $o$-H$_2$O$^+$ and OH$^+$ absorption profiles in velocity space, and agrees with the findings of \citet{neufeld2010}.  The H$_3$O$^+$ transition frequencies are from \citet{yu2009}. Energy level diagrams depicting the specific states studied herein are available for OH$^+$ \citep{lopezsepulcre2013}, H$_2$O$^+$ \citep{schilke2010,ossenkopf2010}, and H$_3$O$^+$ \citep{verhoeve1988,verhoeve1989}.}
\tablenotetext{a}{Indicates the strongest of the hyperfine transitions for a specific $\Delta J$ which was used to set the velocity scale during our analysis.}
\end{deluxetable}
\normalsize

\clearpage
\begin{deluxetable}{lccrrcc}
\tablewidth{0pt}
\tabletypesize{\scriptsize}
\tablecaption{Target List\label{tbl_target}}
\tablehead{\colhead{Target} & \colhead{Right Ascension} & \colhead{Declination} & \colhead{Gal. Long.} & \colhead{Gal. Lat.} & \colhead{Distance} & \colhead{Ref.} \\
 & (hh:mm:ss.s) & (dd:mm:ss.s) & (deg) & (deg) & (kpc) & }
\startdata
M$-$0.13$-$0.08 [SgrA$^*$+20 km~s$^{-1}$ cloud] & 17:45:37.4 & $-$29:05:40.0 & 359.8653 & $-$0.0831 & 8.34 & 1,2 \\
M$-$0.02$-$0.07 [SgrA$^*$+50 km~s$^{-1}$ cloud] & 17:45:50.5 & $-$28:59:53.3 & 359.9724 & $-$0.0737 & 8.34 & 1,2 \\
Sgr B2(M) & 17:47:20.6 & $-$28:23:03.3 & 0.6681 & $-$0.0364 & 8.34 & 1,3,4 \\
Sgr B2(N) & 17:47:20.1 & $-$28:22:18.5 & 0.6778 & $-$0.0284 & 8.34 & 1,3,4 \\
W28A [G005.89$-$00.39] & 18:00:30.5  & $-$24:03:59.9 & 5.8857 & $-$0.3924 & 1.28 & 5 \\
W31C [G010.62$-$00.38] & 18:10:28.7 & $-$19:55:50.0 & 10.6234 & $-$0.3838 & 4.95 & 6 \\
W33A [G012.91$-$00.26] & 18:14:39.2 & $-$17:52:00.5 & 12.9078 & $-$0.2592 & 2.4 & 7 \\
G029.96$-$00.02 & 18:46:03.8 & $-$02:39:22.0 & 29.9556 & $-$0.0163 & 5.26 & 8 \\
\enddata
\tablecomments{Names in brackets are alternate identifiers for the targets.  All coordinates are in the J2000.0 system and indicate where the telescope was pointed.  They may not exactly match published coordinates for the background sources, but in all cases pointings are close enough that the sources are well within the telescope beam (about 19\arcsec, 22\arcsec, and 34\arcsec\ FWHM at 1100~GHz, 950~GHz, and 600~GHz, respectively). Distance references are given below; unless otherwise noted, distance determinations are from trigonometric parallax. 
\newline
References: 1-\citet{reid2014}; 2-\citet{ferriere2012}; 3-\citet{reid2009_sgrb2}; 4-\citet{molinari2011}; 5-\citet{motogi2011}; 6-\citet{sanna2014}; 7-\citet{immer2013}; 8-\citet{zhang2014}; 9-\citet[][kinematic analysis]{fish2003}; 10-\citet{zhang2013}; 11-\citet{sato2010}; 12-\citet{rygl2012}; 13-\citet{moscadelli2009}; 14-\citet{imai2000}; 15-\citet{hachisuka2006}; 16-\citet[][kinematic analysis]{urquhart2012}; 17-\citet{wu2014}}
\end{deluxetable}
\normalsize

\begin{deluxetable}{lccccccc}
\tabletypesize{\scriptsize}
\tablecaption{Double Side Band Continuum Level Antenna Temperature and RMS Noise \label{tbl_obs}} 
\tablehead{\colhead{Source} & \colhead{909 GHz} & \colhead{971 GHz} & \colhead{1033 GHz} &
\colhead{1115 GHz} & \colhead{607 GHz} & \colhead{631 GHz} & \colhead{984 GHz} \\
 & $T_A({\rm DSB})$ & $T_A({\rm DSB})$ & $T_A({\rm DSB})$ &
$T_A({\rm DSB})$ & $T_A({\rm DSB})$ & $T_A({\rm DSB})$ & $T_A({\rm DSB})$ \\
 & (K) & (K) & (K) & (K) & (K) & (K) & (K)}
\startdata
M$-$0.13$-$0.08 & ... & 1.45$\pm$0.07 & ... & 1.40$\pm$0.07 & ... & ... & 1.45$\pm$0.05 \\
M$-$0.02$-$0.07 & ... & 0.85$\pm$0.02 & ... & 0.90$\pm$0.03 & 0.44$\pm$0.02 & 0.51$\pm$0.02 & 0.90$\pm$0.02 \\
Sgr B2(M)\tablenotemark{a} & 17.53$\pm$0.11 & 19.34$\pm$0.31 & 20.63$\pm$0.21 & 22.86$\pm$0.21 & ... & ... & 19.63$\pm$0.22 \\
Sgr B2(N)\tablenotemark{a} & 16.60$\pm$0.12 & 17.86$\pm$0.20 & 17.99$\pm$0.19 & 19.07$\pm$0.38 & ... & ... & 17.86$\pm$0.20 \\
W28A & ... & 3.78$\pm$0.03 & ... & 5.47$\pm$0.04 & 1.01$\pm$0.02 & ... & 3.76$\pm$0.03 \\
W31C & 4.87$\pm$0.04 & 5.46$\pm$0.04 & ... & 6.45$\pm$0.08 & 1.55$\pm$0.01 & 1.75$\pm$0.01 & 5.44$\pm$0.04 \\
W33A & ... & 1.73$\pm$0.02 & ... & 2.18$\pm$0.01 & 0.54$\pm$0.01 & 0.61$\pm$0.01 & 1.72$\pm$0.01 \\
G029.96$-$00.02 & 1.74$\pm$0.01 & 2.08$\pm$0.02 & ... & 2.80$\pm$0.03 & 0.51$\pm$0.01 & ... & 2.11$\pm$0.03 \\
\enddata
\tablecomments{Shown here is the continuum level double sideband antenna temperature and root mean square (RMS) noise resulting from the combination of all observations of each transition.    The full list of ObsIDs used to generate the final spectrum for each transition is given in Table \ref{tbl_obsids}.}
\tablenotetext{a}{These two sight lines were observed in Spectral Scan Mode which produces a single sideband spectrum as the output data product.  To approximate the DSB antenna temperatures we have simply doubled the single sideband antenna temperatures.}
\end{deluxetable}
\normalsize

\clearpage
\begin{deluxetable}{lcrrcccccc}
\tablecolumns{10}
\tablewidth{8.5in}
\rotate
\tabletypesize{\scriptsize}
\tablecaption{Derived column densities \label{tbl_columns}} 
\tablehead{Source & $v_{\rm LSR}$ range & $N({\rm OH}^+)$ & $N({\rm OH}^+)$ & $N({\rm OH}^+)$ & $N(o$-${\rm H_2O^+})$ & $N(p$-${\rm H_2O^+})$ & $N(p$-${\rm H_2O^+})$ & H$_2$O$^+$ OPR & H$_3$O$^+$~$N(1_{0}^{+})$  \\
 &  & 909~GHz & 971~GHz & 1033~GHz & 1115~GHz & 607~GHz & 631~GHz &  & 984~GHz \\
  & (km~s$^{-1}$) & ($10^{13}$ cm$^{-2}$) &  ($10^{13}$ cm$^{-2}$) & ($10^{13}$ cm$^{-2}$) & ($10^{13}$ cm$^{-2}$) & 
($10^{13}$ cm$^{-2}$) & ($10^{13}$ cm$^{-2}$) &  & ($10^{13}$ cm$^{-2}$)
}
\startdata
  M$-$0.13$-$0.08  &                  [-210, -159] &            ... &         $>$7.88 &           ... &   2.40$\pm$1.38 &            ... &           ... &           ... &            ...  \\
  M$-$0.13$-$0.08  &                  [-159, -133] &            ... &        $>$16.88 &           ... &  12.59$\pm$4.94 &            ... &           ... &           ... &            ...  \\
  M$-$0.13$-$0.08  &                   [-133, -92] &            ... &        $>$23.95 &           ... &  12.29$\pm$3.52 &            ... &           ... &           ... &            ...  \\
  M$-$0.13$-$0.08  &                    [-92, -50] &            ... &        $>$22.49 &           ... &   4.16$\pm$1.40 &            ... &           ... &           ... &        $<$2.98  \\
  M$-$0.13$-$0.08  &                    [-50, -39] &            ... &   5.11$\pm$1.42 &           ... &   0.51$\pm$0.29 &            ... &           ... &           ... &        $<$0.78  \\
  M$-$0.13$-$0.08  &                    [-39, -14] &            ... &        $>$10.79 &           ... &   2.12$\pm$0.80 &            ... &           ... &           ... &        $<$1.77  \\
  M$-$0.13$-$0.08  &    [-14, 21]\tablenotemark{a} &            ... &  20.77$\pm$6.20 &           ... &   4.67$\pm$1.39 &            ... &           ... &           ... &  2.40$\pm$2.92  \\
  M$-$0.13$-$0.08  &     [21, 33]\tablenotemark{a} &            ... &   1.96$\pm$0.55 &           ... &   0.45$\pm$0.30 &            ... &           ... &           ... &        $<$0.85  \\
\enddata
\tablecomments{Column densities in the indicated velocity intervals determined from individual transitions of OH$^+$, H$_2$O$^+$, and H$_3$O$^+$ are presented here.  Because the total OH$^+$ and H$_2$O$^+$ column densities for Sgr~B2(M) and Sgr~B2(N) are determined by simultaneously fitting all available transitions, we do not report individual column densities in those sight lines. Also shown are {\it ortho}-to-{\it para} ratios for H$_2$O$^+$ in cases where $p$-H$_2$O$^+$ is detected.  For Sgr~B2(M) {\it ortho}-to-{\it para} ratios are determined using the data from \citet{schilke2013}. Upper limits are given when absorption lines are not detected, and lower limits are given when the absorption features are saturated; both are reported at the $1\sigma$ level.}
\tablenotetext{a}{Denotes velocity range within 5~km~s$^{-1}$ of the background source systemic velocity.}
\end{deluxetable}
\clearpage
\normalsize


\clearpage
\begin{deluxetable}{lccccccccc}
\tablewidth{8.5in}
\rotate
\tabletypesize{\scriptsize}
\tablecaption{Total column densities and inferred results \label{tbl_results}} 
\tablehead{Source & $v_{\rm LSR}$ range & $N({\rm OH}^+)$ & $N({\rm H_2O^+})$ & $N({\rm OH}^+)/N({\rm H_2O^+})$ & $N({\rm H})$ & $R_{gal}$ & $d$ & $f_{\rm H_2}$ & $\zeta_{\rm H}$ \\
  & (km~s$^{-1}$) & ($10^{13}$ cm$^{-2}$) &  ($10^{13}$ cm$^{-2}$) &  & ($10^{21}$ cm$^{-2}$) & (kpc) & (kpc) &  &  ($10^{-16}$~s$^{-1}$)
}
\startdata
  M$-$0.13$-$0.08  &                    [-210, -159] &         $>$7.88 &  3.20$\pm$1.83 &      $\geq$2.46 &        $<$0.84 &          0.0 &           8.3 &        $<$0.121 &                               $>$33.13     \\
  M$-$0.13$-$0.08  &                    [-159, -133] &        $>$16.88 & 16.78$\pm$6.59 &      $\geq$1.01 &  2.29$\pm$0.43 &          0.0 &           8.3 &        $<$0.595 &                              $>$114.09     \\
  M$-$0.13$-$0.08  &                     [-133, -92] &        $>$23.95 & 16.38$\pm$4.70 &      $\geq$1.46 &  1.18$\pm$0.68 &          0.0 &           8.3 &        $<$0.267 &                              $>$147.29     \\
  M$-$0.13$-$0.08  &                      [-92, -50] &        $>$22.49 &  5.55$\pm$1.86 &      $\geq$4.05 &        $>$5.66 &          0.0 &           8.3 &        $<$0.065 &                                    ...     \\
  M$-$0.13$-$0.08  &                      [-50, -39] &   5.11$\pm$1.42 &  0.68$\pm$0.38 &   7.51$\pm$4.73 &  1.91$\pm$0.18 &          3.1 &           5.2 & 0.032$\pm$0.022 &                          3.47$\pm$1.81     \\
  M$-$0.13$-$0.08  &                      [-39, -14] &        $>$10.79 &  2.82$\pm$1.06 &      $\geq$3.83 &  3.24$\pm$0.41 &          4.5 &           3.8 &        $<$0.069 &                                $>$7.46     \\
  M$-$0.13$-$0.08  &      [-14, 21]\tablenotemark{a} &  20.77$\pm$6.20 &  6.23$\pm$1.85 &   3.33$\pm$1.40 &       $>$25.76 &          0.0 &           8.3 & 0.082$\pm$0.043 &                                $<$2.06     \\
  M$-$0.13$-$0.08  &       [21, 33]\tablenotemark{a} &   1.96$\pm$0.55 &  0.60$\pm$0.40 &   3.26$\pm$2.37 &        $>$4.87 &          0.0 &           8.3 & 0.084$\pm$0.076 &                                $<$1.05     \\
\enddata
\tablecomments{References for $N({\rm H})$ are as follows: Winkel et al. 2015 (in prep.)--Sgr~B2(M), Sgr~B2(N), W31C, W33A, G034.3+00.15, W49N, W51e, DR21C, DR21(OH); extracted from the Canadian Galactic Plane Survey \citep[CGPS;][]{taylor2003} data (Winkel, private communication)--AFGL~2591, W3~IRS5, W3(OH); extracted from H~\textsc{i} spectra in \citet{fish2003}--W28A, G029.96$-$00.02, NGC~6334~I, NGC~6334~I(N); extracted from H~\textsc{i} data cube in \citet{lang2010}--M$-$0.02$-$0.07; extracted from H~\textsc{i} optical depth in position 7 of \citet{dwarakanath2004}--M$-$0.13$-$0.08; extracted from H~\textsc{i} spectrum toward G111.61+0.37 in \citet{lebron2001}--NGC~7538~IRS1; extracted from H~\textsc{i} spectum in \citet{urquhart2012}--G327.30$-$00.60. Galactocentric radius and distance are from the kinematic analysis unless otherwise noted in Section \ref{subsec_kind}. Ionization rates are calculated using a scaling factor of $\epsilon=0.07$ \citep{indriolo2012w51}.}
\tablenotetext{a}{Denotes velocity range within 5~km~s$^{-1}$ of the background source systemic velocity.}
\end{deluxetable}
\clearpage
\normalsize

\clearpage
\appendix
\setcounter{table}{0}
\renewcommand{\thetable}{A\arabic{table}}
\clearpage
\begin{deluxetable}{lll}
\tablewidth{8.5in}
\rotate
\tabletypesize{\scriptsize}
\tablecaption{Observation IDs \label{tbl_obsids}}
\tablehead{ \colhead{Source} & \colhead{Transition} & \colhead{Obs IDs}
}
\startdata
M$-$0.13$-$0.08 & OH$^+$ 971~GHz\tablenotemark{a} & 1342228618, 1342228619 \\
M$-$0.13$-$0.08 & H$_2$O$^+$ 1115~GHz & 1342228615, 1342228616 \\
M$-$0.02$-$0.07 & H$_2$O$^+$ 607~GHz & 1342206354, 1342206355, 1342206356 \\
M$-$0.02$-$0.07 & H$_2$O$^+$ 631~GHz & 1342206351, 1342206352, 1342206353 \\
M$-$0.02$-$0.07 & OH$^+$ 971~GHz\tablenotemark{a} & 1342214428, 1342214429, 1342214430 \\
M$-$0.02$-$0.07 & H$_3$O$^+$ 984~GHz & 1342253697 \\
M$-$0.02$-$0.07 & H$_2$O$^+$ 1115~GHz & 1342205882, 1342205883, 1342205884 \\
M$-$0.02$-$0.07 & H$_3$O$^+$ 1655~GHz & 1342216667, 1342216668, 1342216669, 1342216670, 1342216671, 1342216672 \\
\enddata
\tablecomments{Observations were taken as part of the key programs PRISMAS (PRobing InterStellar Molecules with Absoprtion line Studies; PI--Maryvonne Gerin), WISH (Water In Star-Forming regions with {\it Herschel}; PI--Ewine van Dishoeck) and HEXOS ({\it Herschel} observations of EXtra-Ordinary Sources; PI--Ted Bergin) and the open time programs OT1\_dneufeld\_1 (PI--David Neufeld), OT1\_vossenko\_4 (PI--Volker Ossenkopf), and OT1\_cpersson\_1 (PI--Carina Persson).}
\tablenotetext{a}{The H$_3$O$^+$ 0$_0^-$--1$_0^+$ transition at 984 GHz is also covered in this observation.}
\tablenotetext{b}{The OH$^+$ $J'$--$J''=1$--1 transition at 1033~GHz is also covered in this observation.}
\tablenotetext{c}{The $o$-H$_2$O$^+$ $J'$--$J''=1/2$--1/2 transition at 1139~GHz is also covered in this observation.}
\end{deluxetable}
\normalsize
\clearpage

\end{document}